\documentclass[lettersize,journal]{IEEEtran}
\usepackage{amssymb}
\usepackage{algorithm}
\usepackage{algorithmic}
\usepackage{amsmath,amsfonts}
\usepackage{amsthm}
\usepackage{array}
\usepackage{arydshln}
\usepackage{bbding}
\usepackage{bigints}
\usepackage{booktabs}
\usepackage[caption=false,font=normalsize,labelfont=sf,textfont=sf]{subfig}
\usepackage{cite}
\usepackage{color}
\usepackage{diagbox}
\usepackage{epsfig,latexsym}
\usepackage{epstopdf}
\usepackage{graphicx}
\usepackage{fancyhdr}
\usepackage{float}
\usepackage{flushend}
\usepackage{indentfirst}
\usepackage{lastpage}
\usepackage{makecell}
\usepackage{mathtools}
\usepackage{multirow}
\usepackage{pifont}
\usepackage{psfrag}
\usepackage{setspace}
\usepackage{stfloats}
\usepackage{subfig}
\usepackage{subfloat}
\usepackage{textcomp}
\usepackage{times}
\usepackage{verbatim}
\usepackage{url}
\usepackage{hyperref}
\usepackage{balance}
\usepackage{cleveref}
\newcommand{\cmark}{\ding{52}}

\theoremstyle{remark}
\newtheorem{theorem}{\quad \textbf{Theorem}}

\newtheorem{remark}{\quad \textbf{Remark}}

\allowdisplaybreaks[4]

\begin{document}
\title{Stacked Intelligent Metasurface-Based Transceiver Design for Near-Field Wideband Systems}
\author{Qingchao Li, \textit{Member, IEEE},
Mohammed El-Hajjar, \textit{Senior Member, IEEE},\\
Chao Xu, \textit{Senior Member, IEEE},
Jiancheng An, \textit{Member, IEEE},\\
Chau Yuen, \textit{Fellow, IEEE},
and Lajos Hanzo, \textit{Life Fellow, IEEE}

\thanks{This work was supported by the Future Telecoms Research Hub, Platform for Driving Ultimate Connectivity (TITAN), sponsored by the Department of Science Innovation and Technology (DSIT) and the Engineering and Physical Sciences Research Council (EPSRC) under Grant EP/X04047X/1 and Grant EP/Y037243/1. M. El-Hajjar would like to acknowledge the financial support of the EPSRC projects under grant EP/X04047X/2, EP/X04047X/1 and EP/Y037243/1. C. Yuen would like to acknowledge the financial support of the Ministry of Education, Singapore, under its Ministry of Education (MOE) Tier 2 (Award number T2EP50124-0032). L. Hanzo would like to acknowledge the financial support of the EPSRC projects under grant EP/Y037243/1, EP/W016605/1, EP/X01228X/1, EP/Y026721/1, EP/W032635/1, EP/Y037243/1 and EP/X04047X/1 as well as of the European Research Council's Advanced Fellow Grant QuantCom (Grant No. 789028). \textit{(Corresponding author: Lajos Hanzo.)}

Qingchao Li, Mohammed El-Hajjar, Chao Xu and Lajos Hanzo are with the School of Electronics and Computer Science, University of Southampton, Southampton SO17 1BJ, U.K. (e-mail: qingchao.li@soton.ac.uk; meh@ecs.soton.ac.uk; cx1g08@ecs.soton.ac.uk; lh@ecs.soton.ac.uk).

Jiancheng An and Chau Yuen are with the School of Electrical and Electronics Engineering, Nanyang Technological University, Singapore 639798 (e-mail: jiancheng.an@ntu.edu.sg; chau.yuen@ntu.edu.sg).}}

\maketitle

\begin{abstract}
Intelligent metasurfaces may be harnessed for realizing efficient holographic multiple-input and multiple-output (MIMO) systems, at a low hardware-cost and high energy-efficiency. As part of this family, we propose a hybrid beamforming design for stacked intelligent metasurfaces (SIM) aided wideband wireless systems relying on the near-field channel model. Specifically, the holographic beamformer is designed based on configuring the phase shifts in each layer of the SIM for maximizing the sum of the baseband eigen-channel gains of all users. To optimize the SIM phase shifts, we propose a layer-by-layer iterative algorithm for optimizing the phase shifts in each layer alternately. Then, the minimum mean square error (MMSE) transmit precoding method is employed for the digital beamformer to support multi-user access. Furthermore, the mitigation of the SIM phase tuning error is also taken into account in the digital beamformer by exploiting its statistics. The power sharing ratio of each user is designed based on the iterative waterfilling power allocation algorithm. Additionally, our analytical results indicate that the spectral efficiency attained saturates in the high signal-to-noise ratio (SNR) region due to the phase tuning error resulting from the imperfect SIM hardware quality. The simulation results show that the SIM-aided holographic MIMO outperforms the state-of-the-art (SoA) single-layer holographic MIMO in terms of its achievable rate. We further demonstrate that the near-field channel model allows the SIM-based transceiver design to support multiple users, since the spatial resources represented both by the angle domain and the distance domain can be exploited.
\end{abstract}

\section{Introduction}
\IEEEPARstart{N}{ext-generation} wireless systems rely on sophisticated technologies, such as millimeter wave (mmWave) transceivers~\cite{li2020dynamic}, \cite{tang2022path}, massive multiple-input and multiple-output (MIMO) schemes~\cite{li2023uav}, \cite{du2022tensor}, and ultra-dense networking~\cite{wu2020user}. However, their throughput is still insufficient for supporting augmented reality, mixed reality and virtual reality~\cite{torres2020immersive}. These emerging applications require the fusion of interdisciplinary techniques, including communications, sensing, control and edge computing~\cite{liu2024next}. To enhance the efficiency and reliability of next-generation multimedia services, the wireless research community is aiming for harnessing artificial intelligence \cite{zhu2020toward}, Terahertz (THz) techniques~\cite{xu2021graph},\cite{dai2022delay}, and programmable metasurfaces~\cite{wu2021intelligent}.

\subsection{Related Work}
Programmable metasurfaces constitute promising technique as a benefit of their high energy efficiency, low hardware cost, wide coverage and convenient deployment~\cite{you2020channel}. Specifically, a programmable metasurface is composed of novel two-dimensional metamaterials, where a controllable electromagnetic (EM) field can be formed by intelligently reconfiguring the EM waves on the programmable metasurfaces. A typical application of programmable metasurfaces is the popular reconfigurable intelligent surface (RIS)~\cite{li2022reconfigurable}, \cite{li2024low}. Briefly, the RIS is a specific type of programmable metasurface, typically composed of metamaterial unit cells, which are designed to manipulate EM waves in various ways. While reflective elements are commonly studied, the RIS can also support other functionalities, such as absorptive, refractive, and simultaneous transmission and reflection (STAR)~\cite{li2018metasurfaces}. These diverse functionalities make the RIS a versatile tool in optimizing wireless communication systems. The channel environment can be beneficially ameliorated by adjusting the phase shift of the impinging signal on each reconfigurable element. However, the transmitted signal is substantially attenuated by the two-hop path-loss of the transmitter-RIS and the RIS-receiver links~\cite{pan2021reconfigurable}, \cite{pan2022overview}. As a design alternative, the attractive concept of holographic MIMO has been proposed. In contrast to the RIS, which plays the role of a passive relay conceived for `reconfiguring' the propagation environment, the holographic surfaces act as a reconfigurable antenna array at the base stations (BSs). The holographic beamforming architecture relies on a programmable metasurface paradigm in support of improved spectral efficiency and reduced power consumption. This ambitious objective is achieved by harnessing a spatially near-continuous aperture and holographic radios, thus significantly reducing the power consumption and fabrication cost~\cite{huang2020holographic}, \cite{li2024achievable}, \cite{yoo2023sub}, \cite{deng2023reconfigurable}, \cite{gong2024holographic}.

In~\cite{deng2021reconfigurable_tvt}, \cite{deng2022hdma}, \cite{deng2022reconfigurable_twc}, \cite{hu2022holographic}, Deng \textit{et al.} proposed a novel hybrid beamforming architecture relying on a special leaky-wave antenna constituted by a reconfigurable holographic surface (RHS), where the digital beamformer and the holographic beamformer are optimized alternately for maximizing the achievable sum-rate. Specifically, the digital beamformer is designed based on the classical zero-forcing (ZF) precoding method, while the holographic beamformer is optimized based on the control of the amplitude response of the RHS elements. It was demonstrated that the RHS-based hybrid beamformer improves the sum-rate, while reducing the hardware cost, compared to the conventional hybrid digital-analog beamforming architecture relying on phase shifters. In~\cite{deng2022holographic}, an RHS-based beamformer was employed in low-Earth-orbit (LEO) satellite communications for improving the channel's power gain. Furthermore, the sum-rate comparison of the conventional phased array and of an RHS-aided system was presented in~\cite{hu2023holographic}. To reduce the pilot overhead required for channel state information (CSI) acquisition, Wu \textit{et al.}~\cite{wu2024two} proposed a two-timescale beamformer architecture, where the holographic beamformer was optimized based on the statistical CSI, and then the instantaneous CSI of the equivalent channel links was estimated and utilized for the digital beamformer design.

Moreover, the holographic beamformer can also be implemented using a dynamic metasurface antenna (DMA), which comprises multiple microstrips, with each microstrip containing numerous sub-wavelength, frequency-selective resonant metamaterial radiating elements~\cite{shlezinger2019dynamic}, \cite{you2022energy}, \cite{li2023near}. In the DMA, the beamforming design is achieved by linearly combining the radiation observed from all metamaterial elements within each microstrip. The mathematical framework for DMA-based massive MIMO systems was initially proposed by Shlezinger \textit{et al.} in~\cite{shlezinger2019dynamic}, where the fundamental limits of DMA-assisted uplink communications were also explored. In~\cite{you2022energy}, You \textit{et al.} optimized the energy efficiency of the DMA-based massive MIMO system using the Dinkelbach transform, alternating optimization, and deterministic equivalent techniques. Additionally, Li \textit{et al.}~\cite{li2023near} proposed a power-efficient DMA operating at high frequencies, enabling the implementation of extremely large-scale MIMO (XL-MIMO) schemes.

The above beamforming architecture is based on a single-layer metasurface. To further improve both the spatial-domain gain and the beamformer's degree-of-freedom, the authors of~\cite{an2023stacked}, \cite{an2023stacked_icc}, \cite{lin2024stacked} proposed a holographic beamforming paradigm relying on stacked intelligent metasurfaces (SIM) to carry out advanced signal processing directly in the native EM wave regime without a digital beamformer. Specifically, the SIM is composed of stacked reconfigurable multi-layer surfaces, and the phase shifts of the reconfigurable elements found in each layer can be appropriately adjusted for designing the holographic beamformer. This multi-layer architecture not only increases the number of controllable parameters but also allows for hierarchical beamforming, enabling fine-grained manipulation of electromagnetic waves. As a result, the SIM can perform advanced signal processing directly in the native electromagnetic (EM) wave regime, eliminating the need for a digital beamformer and significantly improving the beamforming resolution and flexibility~\cite{an2023stacked}, \cite{an2023stacked_icc}, \cite{lin2024stacked}. In~\cite{an2023stacked}, the gradient descent algorithm was employed for optimizing the SIM phase shifts to maximize the achievable sum-rate, and it was shown that the SIM-based beamforming architecture outperforms its single-layer metasurface based counterparts. In~\cite{an2023stacked_icc}, an alternating optimization method was designed for jointly optimizing the power allocation and SIM-based holographic beamformer in the multi-user multiple-input and single-output (MISO) downlink. Specifically, in each iteration the transmit power allocated to users is based on the classical water-filling algorithm, while the optimization of the SIM phase shift is based on the projected gradient ascent or successive refinement method. Furthermore, in~\cite{lin2024stacked} the SIM technology was leveraged for LEO satellite communication systems. Considering the challenges of acquiring the CSI between the LEO satellite and the ground users, the SIM phase shifts were optimized for maximizing the ergodic sum-rate based on statistical CSI.

\subsection{Motivation}
The above holographic beamforming architecture has the following limitations. Firstly, the existing holographic beamforming designs focus on narrowband signals, hence convey limited data rate. However, practical wireless networks typically utilize wideband signals to achieve higher data rate~\cite{kaiser2009overview}. Secondly, the above holographic beamforming designs are based on the assumption of ideal phase shift control at the reconfigurable elements. But having phase tuning errors is inevitable at the reconfigurable surfaces relying on practical hardware, resulting in significant performance degradation~\cite{badiu2019communication}. Thirdly, the above holographic beamformers are designed based on the far-field channel model. However, for large arrays used for high-frequency communication between extremely large antenna arrays, the near-field range can be as high as tens or even hundreds of meters~\cite{cui2022near}, \cite{an2023toward_beamfocusing}.

Motivated by the above observations, we formulate a SIM-aided transceiver design for near-field wideband systems. Our contributions may be summarized as follows:

\begin{itemize}
  \item We formulate a hybrid beamforming architecture for SIM-aided multi-user wideband wireless systems operating in the face of realistic phase tuning errors. More explicitly, we optimize the holographic beamformer and the digital beamformer to maximize the spectral efficiency. We then solve this non-convex optimization problem by decomposing it into several sub-problems. Firstly, in the holographic beamformer at the SIM, we propose a layer-by-layer iterative optimization algorithm for maximizing the sum of the baseband eigen-channel gains of all users. This alternately optimizes the phase shift matrix of the reconfigurable elements in each layer. Then, the digital beamformer is designed based on the minimum mean square error (MMSE) criterion to mitigate the inter-user interference and the effect of the SIM phase tuning error by exploiting their statistics. This approach differs from traditional narrowband models, as we simultaneously optimize across multiple subcarrier frequencies, rather than individually for all narrowband uses. Finally, the power share of each user is designed based on the iterative waterfilling power allocation algorithm.
  \item To investigate the effect of SIM phase tuning errors, we theoretically derive the spectral efficiency upper bound for the SIM-aided transceiver design in the face of SIM phase tuning errors. The theoretical analysis demonstrates that the spectral efficiency saturates at the high signal-to-noise ratio (SNR) due to the limitation of the phase tuning errors resulting from the imperfect SIM hardware quality.
  \item Our simulation results verify that the SIM-aided holographic beamforming design can promise higher spectral efficiency than that of the state-of-the-art (SoA) single layer metasurface aided holographic beamforming design. Furthermore, our extensive results demonstrate that the near-field channel is capable of supporting multiple users by exploiting the spatial resources in both the angle domain and the distance domain.
\end{itemize}

Finally, Table~\ref{Table_literature} totally and explicitly contrasts our contributions to the literature~\cite{deng2021reconfigurable_tvt,deng2022hdma,deng2022reconfigurable_twc,
hu2022holographic,deng2022holographic,hu2023holographic,wu2024two,shlezinger2019dynamic,
you2022energy,li2023near,an2023stacked,an2023stacked_icc,lin2024stacked} at a glance.

\begin{table*}
\small
\setlength{\tabcolsep}{4pt}
\begin{center}
\caption{Novelty comparison of our paper to the existing metasurface techniques in literature~\cite{deng2021reconfigurable_tvt}, \cite{deng2022hdma}, \cite{deng2022reconfigurable_twc}, \cite{hu2022holographic}, \cite{deng2022holographic}, \cite{hu2023holographic}, \cite{wu2024two}, \cite{shlezinger2019dynamic}, \cite{you2022energy}, \cite{li2023near}, \cite{an2023stacked}, \cite{an2023stacked_icc}, \cite{lin2024stacked}.}
\label{Table_literature}
\begin{tabular}{*{16}{l}}
\hline
     & \makecell[c]{Our paper} & \cite{deng2021reconfigurable_tvt} & \cite{deng2022hdma} & \cite{deng2022reconfigurable_twc} & \cite{hu2022holographic} & \cite{deng2022holographic} & \cite{hu2023holographic} & \cite{wu2024two} & \cite{shlezinger2019dynamic} & \cite{you2022energy} & \cite{li2023near} & \cite{an2023stacked} & \cite{an2023stacked_icc} & \cite{lin2024stacked} \\
\hline
    Beamforming design & \makecell[c]{\cmark} & \makecell[c]{\cmark} & \makecell[c]{\cmark} & \makecell[c]{\cmark} & \makecell[c]{\cmark} & \makecell[c]{\cmark} & \makecell[c]{\cmark} & \makecell[c]{\cmark} & \makecell[c]{\cmark} & \makecell[c]{\cmark} & \makecell[c]{\cmark} &  \makecell[c]{\cmark} & \makecell[c]{\cmark} & \makecell[c]{\cmark} \\
\hdashline
    Multi-user access & \makecell[c]{\cmark} & \makecell[c]{\cmark} & \makecell[c]{\cmark} & \makecell[c]{\cmark} & \makecell[c]{\cmark} & \makecell[c]{\cmark} & \makecell[c]{\cmark} & \makecell[c]{\cmark} & \makecell[c]{\cmark} & \makecell[c]{\cmark} & \makecell[c]{\cmark} &  & \makecell[c]{\cmark} & \makecell[c]{\cmark} \\
\hdashline
    Multi-layer metasurfaces & \makecell[c]{\cmark} &  &  &  &  &  &  & & & &  & \makecell[c]{\cmark} & \makecell[c]{\cmark} & \makecell[c]{\cmark} \\
\hdashline
    Near-field channel model & \makecell[c]{\cmark} &  &  &  &  &  &  & &  &  & \makecell[c]{\cmark} & & &  \\
\hdashline
    Wideband systems & \makecell[c]{\cmark} &  &  &  &  &  &  &  &  &  & & & &  \\
\hdashline
    Hardware impairment mitigation & \makecell[c]{\cmark} &  &  &  &  &  &  &  &  &  & & & & \\
\hline
\end{tabular}
\end{center}
\end{table*}

\subsection{Organization}
The rest of this paper is organized as follows. In Section~\ref{System_Model}, we present the system model, while the hybrid beamforming design is described in Section~\ref{Beamforming_Design}. Our performance analysis is provided in Section~\ref{Performance_Analysis}, followed by our simulation results in Section~\ref{Numerical_and_Simulation_Results}. Finally, we conclude in Section~\ref{Conclusion}.

\subsection{Notations}
Vectors and matrices are denoted by boldface lower and upper case letters, respectively; $\mathbf{A}^{-1}$, $\mathbf{A}^{\text{T}}$ and $\mathbf{A}^{\text{H}}$ represent the inverse, transpose and Hermitian transpose of the matrix $\mathbf{A}$, respectively; $\odot$ represents the Hadamard product operation; $\angle\mathbf{a}$ and $|\mathbf{a}|$ denote the angle and the amplitude of the complex vector $\mathbf{a}$, respectively; $\|\mathbf{a}\|$ denotes the Euclidean norm of the vector $\mathbf{a}$; $\mathbb{C}^{m\times n}$ is the space of $m\times n$ complex-valued matrices; $\mathbf{1}_{N}$ represents the $N\times1$ vector with all elements being 1; $\mathbf{I}_{N}$ represents the $N\times N$ identity matrix; $\mathbf{Diag}\{\mathbf{a}\}$ denotes a diagonal matrix having elements of $\mathbf{a}$ in order; $[\mathbf{a}]_n$ represents the $n$th element in the vector $\mathbf{a}$ and $[\mathbf{A}]_{m,n}$ is the $(m,n)$th element in the matrix $\mathbf{A}$; $\mathbb{E}[\mathbf{x}]$ and $\mathbf{C}_{\mathbf{x}\mathbf{x}}$ represent the mean and the covariance matrix of the vector $\mathbf{x}$, respectively; $\mathcal{CN}(\boldsymbol{\mu},\mathbf{\Sigma})$ is a circularly symmetric complex Gaussian random vector with the mean $\boldsymbol{\mu}$ and the covariance matrix $\mathbf{\Sigma}$; $[a]^{+}$ denotes the maximum value between $a$ and 0.

\section{System Model}\label{System_Model}
In this section, we present the system model of our SIM-aided transceiver designed for wideband information transfer.

\begin{figure}[!t]
    \centering
    \includegraphics[width=3.2in]{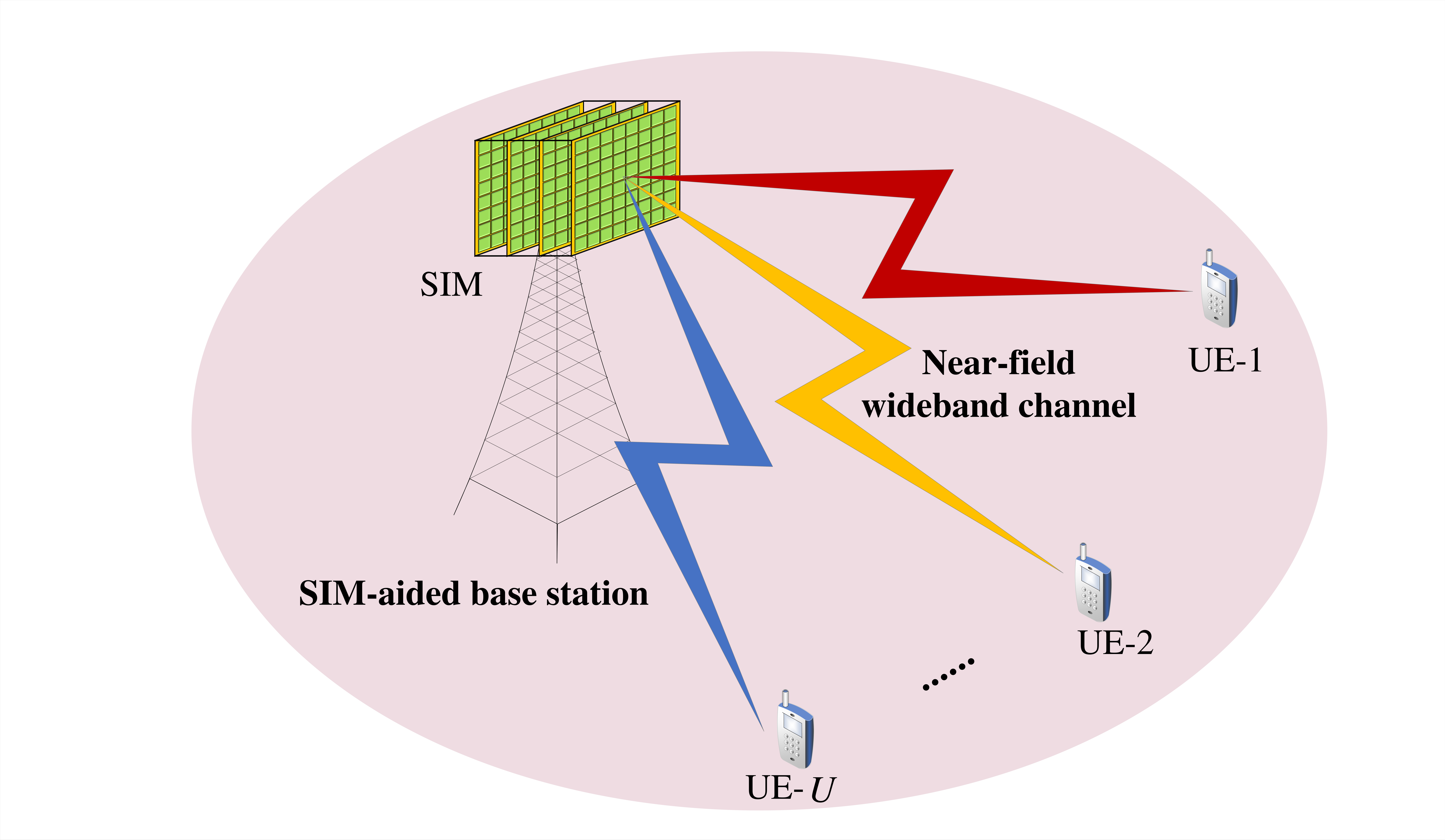}
    \caption{System model of the SIM-aided transceiver design.}\label{Fig_system_model_SIM_BS_UE}
\end{figure}

\begin{figure}[!t]
    \centering
    \includegraphics[width=3.2in]{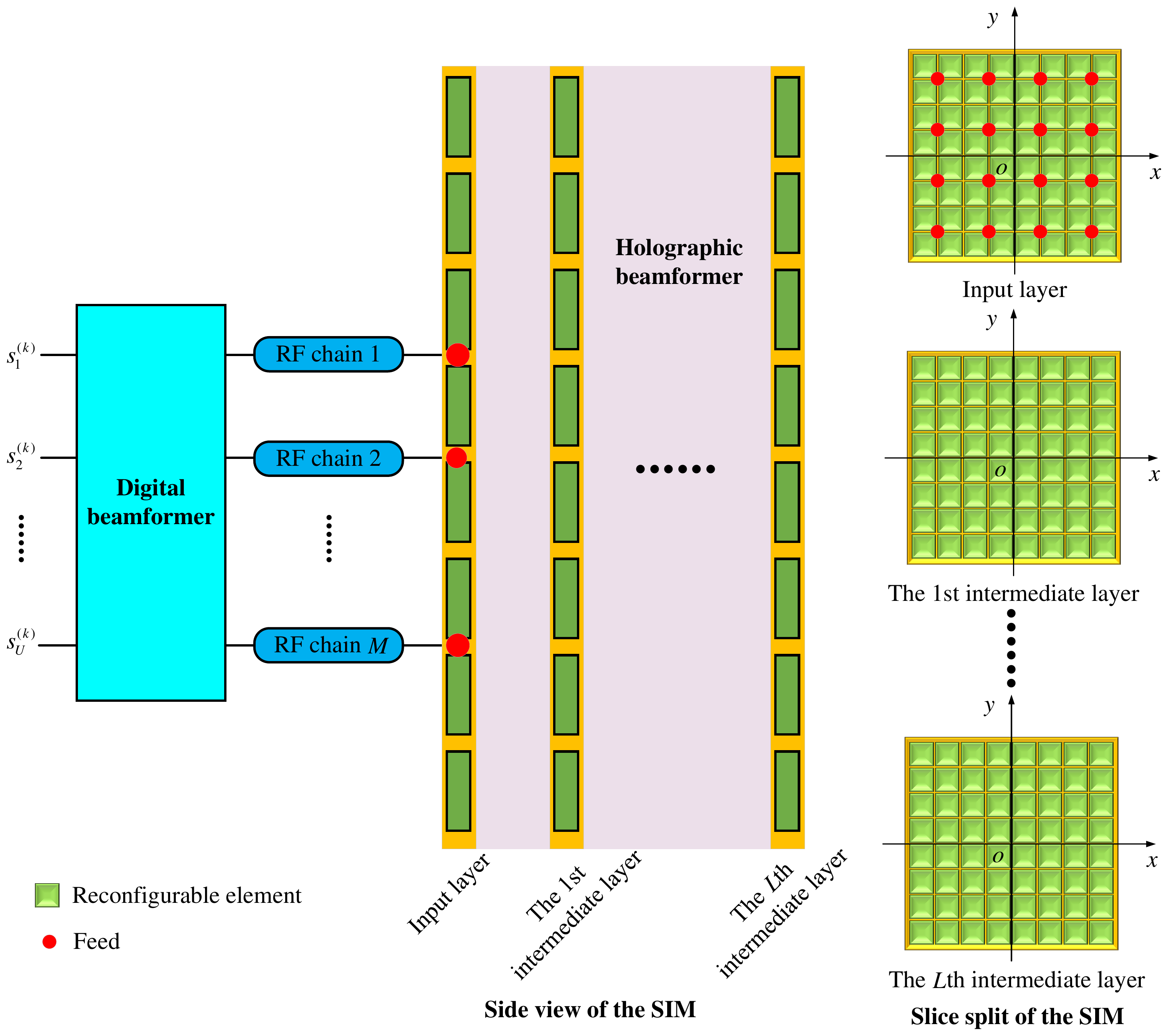}
    \caption{SIM-based hybrid beamforming architecture.}\label{Fig_system_model_SIM_beamforming}
\end{figure}

\subsection{Signal Model}
As shown in Fig. \ref{Fig_system_model_SIM_BS_UE}, without loss of generality we consider the downlink of a wideband wireless system, where a SIM is deployed at the BS to assist transmission from the BS to $U$ single-antenna user equipment (UEs). By harnessing orthogonal frequency-division multiplexing (OFDM), the dispersion of the channel response can be eliminated by partitioning the wide operational bandwidth into low-rate non-dispersive parallel sub-channels. Specifically, we assume that there are $K$ subcarriers in the bandwidth of $B$, with the center carrier frequency of $f_\mathrm{c}$. Furthermore, we denote the transmitted information vector in the $k$th subcarrier as $\mathbf{s}^{(k)}=[s_1^{(k)},s_2^{(k)},\cdots,s_U^{(k)}]^\mathrm{T}$, where $s_u^{(k)}$ is the information transmitted to UE-$u$ ($u=1,2,\cdots,U$) at the carrier frequency of $f_k=f_\mathrm{c}+\frac{B}{K}(k-\frac{K+1}{2})$.

To realize spectral- and energy-efficient information transfer, we propose a hybrid beamforming architecture based on the SIM shown in Fig. \ref{Fig_system_model_SIM_beamforming}. Specifically, in the $k$th subcarrier, the baseband signal $\mathbf{s}^{(k)}\in\mathbb{C}^{U\times1}$ is transmit precoded with the aid of a digital beamformer $\mathbf{V}^{(k)}\in\mathbb{C}^{M\times U}$, and then further processed by the SIM-aided holographic beamformer $\mathbf{A}^{(k)}\in\mathbb{C}^{N\times M}$. Therefore, the signal received at UE-$u$ over the $k$th subcarrier, denoted as $y_u^{(k)}\in\mathbb{C}^{1\times1}$, is given by
\begin{align}\label{Signal_Model_1}
    y_u^{(k)}=\sqrt{\rho_u^{(k)}}\mathbf{g}_u^{(k)\mathrm{H}}\mathbf{A}^{(k)}
    \mathbf{V}^{(k)}\mathbf{s}^{(k)}+w_u^{(k)},
\end{align}
where $\mathbf{g}_u^{(k)\mathrm{H}}\in\mathbb{C}^{1\times N}$ represents the channel spanning from the BS to the UE-$u$, $w_u^{(k)}\sim\mathcal{CN}(0,\sigma_w^2)$ is the additive noise with the noise density of $\sigma_w^2$, $M$ is the number of radio frequency (RF) chains, and $N$ is the number of reconfigurable elements in each SIM layer. Finally, $\rho_u^{(k)}$ is the transmit power allocated to UE-$u$ over the $k$th subcarrier. We further denote the total transmit power over the $k$th subcarrier as $\rho^{(k)}=\sum_{u=1}^{U}\rho_u^{(k)}$.

\subsection{Stacked Intelligent Metasurface Architecture}
As shown in Fig. \ref{Fig_system_model_SIM_BS_UE}, the SIM is constituted by a partially sealed structure having multiple layers of stacked reconfigurable metasurfaces. The structure of a SIM includes an input layer and $L$ intermediate layers, each of which can be intelligently configured for favourable signal propagation, as described below.
\begin{itemize}
  \item Firstly, the input layer consists of three components, including feeds, sub-wavelength metamaterial radiation elements and a waveguide. Specifically, the feeds are embedded into the input layer to generate incident EM waves, as shown in Fig. \ref{Fig_system_model_SIM_beamforming}. In our SIM-aided beamforming architecture, each RF chain is connected to a feed of the SIM and sends the up-converted signals to its connected feed. The feed then transforms the high frequency current into an EM wave, propagating along the surface of the SIM and exciting the EM field of the SIM. The reconfigurable radiation elements are made of artificial composite material, which are capable of adjusting the coefficients of the EM waves with the aid of a software controller, such as a field programmable gate array (FPGA)~\cite{an2023stacked}, \cite{an2023stacked_icc}. Flexible printed circuit boards (PCBs) or coaxial cables are used to connect the FPGA to the reconfigurable radiation elements, ensuring seamless signal transmission, while maintaining the structural integrity of the SIM~\cite{xiao2024multichannel}, \cite{li2023light}. The waveguide plays the role of a guiding wave structure and it is the propagation medium of the EM wave from the feeds to the reconfigurable radiation elements.
  \item The intermediate layers are designed based on reconfigurable refractive surfaces (RRSs), each of which is composed of numerous refractive elements as shown in Fig. \ref{Fig_system_model_SIM_beamforming}. The incident EM waves can propagate through the hole of each RRS layer~\cite{liu2022programmable}. Furthermore, the coefficients of the refractive elements can be adjusted by a software controller.
\end{itemize}

As shown in Fig. \ref{Fig_system_model_SIM_beamforming}, the input layer has $M=M_x\times M_y$ uniform rectangular planar array (URPA) feeds, where $M_x$ and $M_y$ are the number of feeds in the $x$-direction and the $y$-direction, respectively. Furthermore, each metasurface layer has $N=N_x\times N_y$ URPA having densely-spaced reconfigurable radiation/refractive elements, where $N_x$ and $N_y$ are the number of elements in the $x$-direction and the $y$-direction respectively. We denote the distance between the adjacent feeds as $z_x$ and $z_y$ in the $x$-direction and the $y$-direction, respectively, and the physical size of each reconfigurable element as $\delta=\delta_x\times\delta_y$. Furthermore, we denote the distance between the input layer and the 1st intermediate layer as $d_1$, while that between the $l$th intermediate layer and the $(l+1)$st intermediate layer as $d_{l+1}$. As shown in Fig. \ref{Fig_system_model_SIM_beamforming}, the coordinate of the geometrical center of the $m$th feed in the input layer is denoted as $\mathbf{q}_m$. The coordinate of the geometrical center of the $n$th reconfigurable radiation element in the input layer can be represented as $\mathbf{p}_n^{(0)}=(x_n^{(0)},y_n^{(0)},0)^\mathrm{T}$. Moreover, the coordinate of the geometrical center of the $n$th reconfigurable refractive element in the $l$th intermediate layer can be represented as $\mathbf{p}_n^{(l)}=(x_n^{(l)},y_n^{(l)},\sum_{l'=1}^{l}d_l)^\mathrm{T}$.

For ease of expression, we denote the input layer as layer-0, and the $l$th intermediate layer as layer-$l$. In layer-0, we denote the channel matrix between the feeds and the reconfigurable radiation elements over the $k$th subcarrier as $\mathbf{F}_k^{(0)}\in\mathbb{C}^{N\times M}$, where the link spanning from the $m$th feed to the $n$th reconfigurable radiation element can be represented as:
\begin{align}\label{System_Model_1}
    \left[\mathbf{F}_k^{(0)}\right]_{n,m}=\frac{1}{\sqrt{N}}\mathrm{e}^{-\jmath \frac{2\pi}{\lambda_k}
    \left\|\mathbf{p}_{n}^{(0)}-\mathbf{q}_{m}\right\|},
\end{align}
and $\lambda_k=\frac{c}{f_k}$ with $c$ being the speed of light. For $l=1,2,\cdots,L$, we denote the channel matrix from the reconfigurable elements of layer-$(l-1)$ to that of layer-$l$ over the $k$th subcarrier as $\mathbf{F}_k^{(l)}\in\mathbb{C}^{N\times N}$, where the channel link spanning from the $n_1$th element in layer-$(l-1)$ to the $n_2$th element in layer-$l$ can be formulated as:
\begin{align}\label{System_Model_2}
    \left[\mathbf{F}_k^{(l)}\right]_{n_2,n_1}=\sqrt{\beta_{n_2,n_1}^{(l)}}\mathrm{e}^{-\jmath \frac{2\pi}{\lambda_k}
    \left\|\mathbf{p}_{n_2}^{(l)}-\mathbf{p}_{n_1}^{(l-1)}\right\|}.
\end{align}
In (\ref{System_Model_2}) $\beta_{n_2,n_1}^{(l)}$ is the channel's power gain spanning from the $n_1$th element in layer-$(l-1)$ to the $n_2$th element in the layer-$l$, which is given by
\begin{align}\label{System_Model_2_2}
    \notag\beta_{n_2,n_1}^{(l)}
    =&\iiint_{\mathcal{D}_{n_1}^{(l-1)}}\frac{\sin\psi_{\mathbf{p}_{n_2}^{(l)}
    -\mathbf{t}}}{4\pi\|\mathbf{p}_{n_2}^{(l)}-\mathbf{t}\|^2}\mathrm{d}\mathbf{t}\\
    \overset{(a)}=&\iiint_{\mathcal{D}_{n_1}^{(l-1)}}\frac{\|\mathbf{p}_{n_2}^{(l)}\|\cdot
    \sin\psi_{\mathbf{p}_{n_2}^{(l)}}}{4\pi\|\mathbf{p}_{n_2}^{(l)}-\mathbf{t}\|^3}
    \mathrm{d}\mathbf{t}.
\end{align}
In (\ref{System_Model_2_2}) the integration interval of the $n$th element in layer-$l$ is represented as $\notag{\mathcal{D}_{n}^{(l)}}
=\{(x,y,0)^\mathrm{T}:x_{n}^{(l)}-\frac{\delta_x}{2}<x\leq x_{n}^{(l)}+\frac{\delta_x}{2},y_{n}^{(l)}-\frac{\delta_y}{2}<y\leq y_{n}^{(l)}+\frac{\delta_y}{2}\}$, while $\psi_{\mathbf{p}_{n_2}^{(l)}-\mathbf{t}}$ denotes the angle between the vector $\mathbf{p}_{n_2}^{(l)}-\mathbf{t}$ and the $xoy$ plane of Fig. \ref{Fig_system_model_SIM_beamforming}. Furthermore, $\psi_{\mathbf{p}_{n_2}^{(l)}}$ denotes the angle between the vector $\mathbf{p}_{n_2}^{(l)}$ and the $xy$ plane in the Cartesian coordinate system, and the equality in (a) is based on $\|\mathbf{p}_{n_2}^{(l)}-\mathbf{t}\|\cdot\sin\psi_{\mathbf{p}_{n_2}^{(l)}-\mathbf{t}}
=\|\mathbf{p}_{n_2}^{(l)}\|\cdot\sin\psi_{\mathbf{p}_{n_2}^{(l)}}$ for all $\mathbf{t}\in{\mathcal{D}_{n_1}^{(l-1)}}$.

We denote the coefficient of the $n$th element in layer-$l$ by $\Gamma_n^{(l)}\mathrm{e}^{\jmath\theta_n^{(l)}}$, where $\Gamma_n^{(l)}\in[0,1]$ and $\theta_n^{(l)}\in(-\pi,\pi]$ represent the appropriately configured amplitude and phase shift of the $n$th element, respectively. We set $\Gamma_n^{(l)}=1$ to represent perfect energy transfer. Hence, the response of layer-$l$ is given by
\begin{align}\label{System_Model_5}
    \notag\mathbf{\Theta}^{(l)}=&\mathbf{Diag}\left\{\mathrm{e}^{\jmath\theta_1^{(l)}},
    \mathrm{e}^{\jmath\theta_2^{(l)}},\cdots,\mathrm{e}^{\jmath\theta_N^{(l)}}\right\}\\
    =&\mathbf{Diag}\left\{\mathrm{e}^{\jmath\left(\overline{\theta}_1^{(l)}
    +\widetilde{\theta}_1^{(l)}\right)},
    \mathrm{e}^{\jmath\left(\overline{\theta}_2^{(l)}+\widetilde{\theta}_2^{(l)}\right)},\cdots,
    \mathrm{e}^{\jmath\left(\overline{\theta}_N^{(l)}+\widetilde{\theta}_N^{(l)}\right)}\right\},
\end{align}
where $\theta_n^{(l)}=\overline{\theta}_n^{(l)}+\widetilde{\theta}_n^{(l)}$ with $\overline{\theta}_n^{(l)}$ being the expected phase shift configuration of the $n$th element in layer-$l$, while $\widetilde{\theta}_n^{(l)}$ represents the phase tuning errors due to the realistic SIM hardware imperfection. The phase tuning error $\widetilde{\theta}_n^{(l)}$ obeys identically and independently distributed (i.i.d.) random variables having the mean of 0, and it may also be modelled by the von-Mises distribution or the uniform distribution~\cite{badiu2019communication}, \cite{qian2020beamforming}. These may be represented as $\widetilde{\theta}_n^{(l)}\sim\mathcal{VM}(0,\varpi_\text{p})$ and $\widetilde{\theta}_n^{(l)}\sim\mathcal{UF}(-\iota_\text{p},\iota_\text{p})$, respectively, where $\varpi_\text{p}$ is the concentration parameter of the von-Mises distributed variables and $(-\iota_\text{p},\iota_\text{p})$ is the support interval of the uniformly distributed variables. Although the exact values of $\widetilde{\theta}_n^{(l)}$ cannot be obtained, we can exploit its statistics for beamforming designs. The statistical distribution of the phase tuning error can be practically obtained through multiple approaches. Firstly, hardware-specific models, such as those outlined in~\cite{badiu2019communication}, \cite{qian2020beamforming}, provide initial approximations based on the characteristics of the oscillators and phase shifters. Secondly, empirical measurements during hardware testing or system deployment can validate these models and refine the phase tuning error parameters. Lastly, dynamic estimation methods, such as pilot-based channel training~\cite{papazafeiropoulos2021intelligent}, enable real-time tracking of the phase tuning error during operation, allowing adaptive beamforming adjustments. The above approaches ensure that the phase tuning error statistics are both practically measurable and usable for robust beamforming design. For layer-$l$, we represent the desired phase shift matrix and the phase tuning error matrix as $\overline{\mathbf{\Theta}}^{(l)}=\mathbf{Diag}\{\mathrm{e}^{\jmath\overline{\theta}_1^{(l)}},
\mathrm{e}^{\jmath\overline{\theta}_2^{(l)}},\cdots,\mathrm{e}^{\jmath\overline{\theta}_N^{(l)}}\}$ and $\widetilde{\mathbf{\Theta}}^{(l)}=\mathbf{Diag}
\{\mathrm{e}^{\jmath\widetilde{\theta}_1^{(l)}},\mathrm{e}^{\jmath\widetilde{\theta}_2^{(l)}},
\cdots,\mathrm{e}^{\jmath\widetilde{\theta}_N^{(l)}}\}$, respectively. Then we can get
\begin{align}\label{System_Model_6_3}
    \mathbf{\Theta}^{(l)}=\overline{\mathbf{\Theta}}^{(l)}\odot\widetilde{\mathbf{\Theta}}^{(l)}.
\end{align}

Therefore, the equivalent SIM-based holographic beamformer over the $k$th subcarrier is given by
\begin{align}\label{System_Model_7}
    \mathbf{A}^{(k)}=\mathbf{\Theta}^{(L)}\mathbf{F}_k^{(L)}\mathbf{\Theta}^{(L-1)}
    \mathbf{F}_k^{(L-1)}\cdots\mathbf{\Theta}^{(1)}\mathbf{F}_k^{(1)}
    \mathbf{\Theta}^{(0)}\mathbf{F}_k^{(0)}.
\end{align}

\subsection{Channel Model}
When the communication distance between transceivers is shorter than the Rayleigh distance, which is formulated by $\frac{2D^2}{\lambda_\mathrm{c}}$ with $\lambda_\mathrm{c}=\frac{c}{f_\mathrm{c}}$ being the wavelength and $D$ represents the maximum physical dimension of the SIM, the EM waves radiated from the transmitter must be modeled as spherical waves rather than plane waves~\cite{cui2022near}. Given the short range in high-frequency communications, as well as the large number of reconfigurable elements in the metasurface, we employ the near-field model for accurately characterizing the channel response between the BS and the UEs~\cite{an2023toward_beamfocusing}.

We employ the non-uniform spherical wave (NUSW) model for accurately characterizing the near-field channel response between the SIM and the UE, as proposed in~\cite{lu2021communicating}. This model is particularly suitable for the radiative near-field region, where the distance between the SIM and UEs satisfies $r>0.62\sqrt{D^3/\lambda_\mathrm{c}}$. In the radiative near-field region, the EM waves propagate as spherical waves with radiative components dominating, while the reactive components are negligible. Thus, the channel response spanning from the outermost intermediate layer, i.e., layer-$L$, to UE-$u$ over the $k$th subcarrier can be expressed as:
\begin{align}\label{System_Model_8}
    \notag\mathbf{g}_u^{(k)\mathrm{H}}=&\left[
    \sqrt{\zeta_{1}}\mathrm{e}^{-\jmath\frac{2\pi}{\lambda_k}
    \left\|\mathbf{r}_u-\mathbf{p}_{1}^{(L)}\right\|},
    \sqrt{\zeta_{2}}\mathrm{e}^{-\jmath\frac{2\pi}{\lambda_k}
    \left\|\mathbf{r}_u-\mathbf{p}_{2}^{(L)}\right\|},\right.\\
    &\left.\cdots,\sqrt{\zeta_{N}}\mathrm{e}^{-\jmath\frac{2\pi}{\lambda_k}
    \left\|\mathbf{r}_u-\mathbf{p}_{N}^{(L)}\right\|}
    \right],
\end{align}
where $\mathbf{r}_u$ represents the coordinate of UE-$u$, and $\zeta_{n}$ is the channel's power gain spanning from the $n$th element in layer-$L$ to the UE-$u$, given by $\zeta_{n}=\iiint_{\mathcal{D}_{n}^{(L)}}\frac{\left\|\mathbf{r}_u\right\|
\cdot\sin\psi_{\mathbf{r}_u}}{4\pi\left\|\mathbf{r}_u-\mathbf{t}\right\|^3}\mathrm{d}\mathbf{t}$
\footnote{The near-field channel model adopted follows the radiative near-field assumption, which is widely employed in practical holographic MIMO systems~\cite{lu2021communicating}. The reactive near-field components, which are non-trivial only at sub-wavelength distances, are omitted as their impact is negligible in typical wireless communication scenarios~\cite{an2023toward_beamfocusing}. Additionally, we neglect higher-order electromagnetic interactions and phase delay terms for tractability, as commonly done in near-field beamforming studies~\cite{lu2021communicating}. Future extensions could explore more detailed formulations that incorporate reactive field effects and Maxwell-equation-based modeling for extremely large-scale holographic MIMO systems.}. In this formula $\psi_{\mathbf{r}_u}$ denotes the angle between the vector $\mathbf{r}_u$ and the $xy$ plane in the Cartesian coordinate. In practical systems, since the size of each reconfigurable element is on the wavelength scale, the channel's power gain variation from different points belonging to $\mathcal{D}_{n}^{(L)}$ to the UEs is negligible. Therefore, the channel's power gain spanning from the $n$th element in layer-$L$ to the UE-$u$ can be approximated as $\zeta_{n}\approx\frac{\delta\|\mathbf{r}_u\|\cdot\sin\psi_{\mathbf{r}_u}}
{4\pi\|\mathbf{r}_u-\mathbf{p}_n^{(L)}\|^3}$.

\section{Hybrid Digital and Holographic Design}\label{Beamforming_Design}
In this section, we design the digital beamformers and the SIM-based holographic beamformers for maximizing the spectral efficiency of the wideband system.

According to (\ref{System_Model_7}) and (\ref{System_Model_8}), the baseband equivalent channel $\mathbf{h}_u^{(k)\mathrm{H}}\in\mathbb{C}^{1\times M}$ spanning from the RF chains of the BS to UE-$u$ over the $k$th subcarrier, can be expressed as
\begin{align}\label{Beamforming_Design_1}
    \mathbf{h}_u^{(k)\mathrm{H}}=\mathbf{g}_u^{(k)\mathrm{H}}\mathbf{A}^{(k)}.
\end{align}
Note that the baseband equivalent channel $\mathbf{h}_u^{(k)\mathrm{H}}$ is impaired due to the phase tuning error of the reconfigurable metasurface, and only its statistics are known. Specifically, the mean of $\mathbf{h}_k^{\mathrm{H}}$, denoted as $\overline{\mathbf{h}}_u^{(k)\mathrm{H}}$, can be acquired by relying on the statistics of the phase tuning error of the SIM. Therefore, the deterministic baseband equivalent channel is
\begin{align}\label{Beamforming_Design_2}
    \overline{\mathbf{h}}_u^{(k)\mathrm{H}}=\mathbb{E}\left[\mathbf{h}_u^{(k)\mathrm{H}}\right],
\end{align}
and the baseband equivalent channel's uncertainty is
\begin{align}\label{Beamforming_Design_3}
    \widetilde{\mathbf{h}}_u^{(k)\mathrm{H}}=\mathbf{h}_u^{(k)\mathrm{H}}-
    \mathbb{E}\left[\mathbf{h}_u^{(k)\mathrm{H}}\right].
\end{align}

According to (\ref{Signal_Model_1}), (\ref{Beamforming_Design_1}), (\ref{Beamforming_Design_2}) and (\ref{Beamforming_Design_3}), the signal received at UE-$u$ over the $k$th subcarrier can be formulated as
\begin{align}\label{Beamforming_Design_4}
    \notag y_u^{(k)}=&\sqrt{\rho_u^{(k)}}\mathbf{g}_u^{(k)\mathrm{H}}\mathbf{A}^{(k)}
    \mathbf{S}^{(k)}\mathbf{V}^{(k)}\mathbf{s}^{(k)}+w_u^{(k)},\\
    \notag=&\underbrace{\sqrt{\rho_u^{(k)}}\overline{\mathbf{h}}_u^{(k)\mathrm{H}}
    \mathbf{S}^{(k)}\mathbf{v}_u^{(k)}s_u^{(k)}}_{\text{Signal over determinate channel}}
    +\underbrace{\sqrt{\rho_u^{(k)}}\widetilde{\mathbf{h}}_u^{(k)\mathrm{H}}
    \mathbf{S}^{(k)}\mathbf{v}_u^{(k)}s_u^{(k)}}_{\text{Signal over uncertain channel}}\\
    &+\underbrace{\sum_{u'=1,u'\neq u}^{U}\sqrt{\rho_u^{(k)}}\mathbf{h}_u^{(k)\mathrm{H}}
    \mathbf{S}^{(k)}\mathbf{v}_{u'}^{(k)}s_{u'}^{(k)}}_{\text{Inter-user interference}}
    +\underbrace{w_u^{(k)}}_{\text{Additive noise}},
\end{align}
where $\mathbf{v}_u^{(k)}$ represents the $u$th column of the beamforming matrix $\mathbf{V}^{(k)}$. Furthermore, $\mathbf{S}^{(k)}\in\mathbb{C}^{M\times M}$ represents the mutual coupling matrix of the RF chain ports. Specifically, the mutual coupling matrix $\mathbf{S}^{(k)}$ adopting the Z-parameter based model is given by~\cite{chen2018review}
\begin{align}\label{Mutual_couple_1}
    \mathbf{S}^{(k)}=\left(Z_A+Z_L\right)\left(\mathbf{Z}^{(k)}+Z_L\mathbf{I}_M\right)^{-1},
\end{align}
where $Z_A$ is the antenna impedance and $Z_L$ is the load impedance, both of which are fixed as 50 Ohms. Moreover, $\mathbf{Z}^{(k)}\in\mathbb{C}^{M\times M}$ is the mutual impedance matrix, with the $(m_1,m_2)$th entry represented as $[\mathbf{Z}^{(k)}]_{m_1,m_2}=Z_A$ for $m_1=m_2$, and $[\mathbf{Z}^{(k)}]_{m_1,m_2}=60\mathcal{C}_\mathrm{I}(\frac{2\pi d_{m_1,m_2}}{\lambda_k})
-60\mathcal{S}_\mathrm{I}(\frac{2\pi d_{m_1,m_2}}{\lambda_k})-30\mathcal{C}_\mathrm{I}
(\frac{2\pi d_{m_1,m_2}'}{\lambda_k})+30\mathcal{S}_\mathrm{I}
(\frac{2\pi d_{m_1,m_2}'}{\lambda_k})-30\mathcal{C}_\mathrm{I}
(\frac{2\pi d_{m_1,m_2}''}{\lambda_k})+30\mathcal{S}_\mathrm{I}
(\frac{2\pi d_{m_1,m_2}''}{\lambda_k})$ for $m_1\neq m_2$, where $\delta_0$ is the dipole length, $d_{m_1,m_2}$ denotes the distance between the $m_1$th and the $m_2$th RF chain ports, $d_{m_1,m_2}'=\sqrt{d_{m_1,m_2}^2+\delta_0^2}+\delta_0$, $d_{m_1,m_2}''=\sqrt{d_{m_1,m_2}^2+\delta_0^2}-\delta_0$, $\mathcal{C}_\mathrm{I}(\cdot)$ and $\mathcal{S}_\mathrm{I}(\cdot)$ denote the cosine integral and the sine integral, respectively~\cite{balanis2015antenna}.

\begin{theorem}\label{Theorem_1}
    The mean of the baseband equivalent channel $\mathbf{h}_u^{(k)\mathrm{H}}$ is given by
    \begin{align}\label{Beamforming_Design_5}
        \overline{\mathbf{h}}_u^{(k)\mathrm{H}}
        =\xi^{L+1}\mathbf{g}_u^{(k)\mathrm{H}}\overline{\mathbf{A}}^{(k)},
    \end{align}
    and the covariance matrix of $\widetilde{\mathbf{h}}_u^{(k)\mathrm{H}}$ is formulated as:
    \begin{align}\label{Beamforming_Design_6}
        \mathbf{C}_{\widetilde{\mathbf{h}}_u^{(k)}\widetilde{\mathbf{h}}_u^{(k)}}
        =\mathbf{F}_k^{(0)\mathrm{H}}\mathbf{\Phi}_k^{(0)}\mathbf{F}_k^{(0)}
        -\xi^{2(L+1)}\overline{\mathbf{A}}^{(k)\mathrm{H}}\mathbf{g}_u^{(k)}
        \mathbf{g}_u^{(k)\mathrm{H}}\overline{\mathbf{A}}^{(k)}.
    \end{align}
    In (\ref{Beamforming_Design_5}) and (\ref{Beamforming_Design_6}), $\overline{\mathbf{A}}^{(k)}$ is formulated as
    \begin{align}\label{System_Model_6_2}
        \overline{\mathbf{A}}^{(k)}=\overline{\mathbf{\Theta}}^{(L)}\mathbf{F}_k^{(L)}
        \overline{\mathbf{\Theta}}^{(L-1)}\mathbf{F}_k^{(L-1)}\cdots
        \overline{\mathbf{\Theta}}^{(1)}\mathbf{F}_k^{(1)}
        \overline{\mathbf{\Theta}}^{(0)}\mathbf{F}_k^{(0)}.
    \end{align}
    Furthermore, $\mathbf{\Phi}_k^{(l)}=\xi^2\overline{\mathbf{\Theta}}^{(l)\mathrm{H}}
    \mathbf{F}_k^{(l+1)\mathrm{H}}\mathbf{\Phi}_k^{(l+1)}\mathbf{F}_k^{(l+1)}
    \overline{\mathbf{\Theta}}^{(l)}+(1-\xi^2)(\mathbf{F}_k^{(l+1)\mathrm{H}}
    \mathbf{\Phi}_k^{(l+1)}\mathbf{F}_k^{(l+1)})\odot\mathbf{I}_N$ for $l=0,1,\cdots,L-1$, and $\mathbf{\Phi}_k^{(L)}=\xi^2\overline{\mathbf{\Theta}}^{(L)\mathrm{H}}
    \mathbf{g}_u^{(k)}\mathbf{g}_u^{(k)\mathrm{H}}\overline{\mathbf{\Theta}}^{(L)}
    +(1-\xi^2)(\mathbf{g}_u^{(k)}\mathbf{g}_u^{(k)\mathrm{H}})\odot\mathbf{I}_N$.

    In (\ref{Beamforming_Design_4}), (\ref{Beamforming_Design_5}) and (\ref{Beamforming_Design_6}), we have $\xi=\frac{\sin(\iota_\mathrm{p})}{\iota_\mathrm{p}}$ when the SIM phase tuning error follows $\mathcal{U}(-\iota_\mathrm{p},\iota_\mathrm{p})$, and $\xi=\frac{I_1(\varpi_\text{p})}{I_0(\varpi_\mathrm{p})}$ when it obeys $\mathcal{VM}(0,\varpi_\text{p})$ with $I_0(\cdot)$ and $I_1(\cdot)$ representing the modified Bessel functions of the first kind of order 0 and order 1, respectively. The SIM phase tuning error variance is $\sigma_\mathrm{p}^2=
    \mathbb{E}[\widetilde{\theta}_n^{(l)2}]=\frac{1}{3}\iota_\mathrm{p}^2$ and $\sigma_\mathrm{p}^2=\mathbb{E}[\widetilde{\theta}_n^{(l)2}]=\frac{1}{\varpi_\text{p}}$, when it follows the uniform distribution and the von-Mises distribution, respectively.
\end{theorem}
\begin{IEEEproof}
    See Appendix \ref{Appendix_A}.
\end{IEEEproof}

According to (\ref{Beamforming_Design_4}), (\ref{Beamforming_Design_5}) and (\ref{Beamforming_Design_6}), the spectral efficiency of UE-$u$ over the $k$th subcarrier can be formulated as
\begin{align}\label{Beamforming_Design_7}
    R_u^{(k)}=&\log_2\left(1+\frac{\rho_u^{(k)}\left\|\overline{\mathbf{h}}_u^{(k)\mathrm{H}}
    \mathbf{S}^{(k)}\mathbf{v}_u^{(k)}\right\|^2}
    {\mathbf{v}_u^{(k)\mathrm{H}}\mathbf{Q}_u^{(k)}\mathbf{v}_u^{(k)}+\sigma_w^2}\right),
\end{align}
where $\mathbf{Q}_u^{(k)}$ is given by
\begin{align}\label{Beamforming_Design_7_2}
    \notag\mathbf{Q}_u^{(k)}=&\rho_u^{(k)}\mathbf{S}^{(k)\mathrm{H}}\mathbf{C}_{\widetilde{\mathbf{h}}_u^{(k)}
    \widetilde{\mathbf{h}}_u^{(k)}}\mathbf{S}^{(k)}\\
    &+\sum_{u'=1,u'\neq u}^{U}\rho_{u'}^{(k)}
    \mathbf{S}^{(k)\mathrm{H}}
    \left(\overline{\mathbf{h}}_{u'}^{(k)}\overline{\mathbf{h}}_{u'}^{(k)\mathrm{H}}
    +\mathbf{C}_{\widetilde{\mathbf{h}}_{u'}^{(k)}\widetilde{\mathbf{h}}_{u'}^{(k)}}\right)
    \mathbf{S}^{(k)}.
\end{align}
We aim for optimizing the precoding matrices $\mathbf{V}^{(1)},\mathbf{V}^{(2)},\cdots,\mathbf{V}^{(K)}$ and the SIM phase shift designs  $\overline{\mathbf{\Theta}}^{(0)},\overline{\mathbf{\Theta}}^{(1)},\cdots,
\overline{\mathbf{\Theta}}^{(L)}$ to maximize the average spectral efficiency of the wideband system\footnote{In this paper, we focus on maximizing the average spectral efficiency for all users. It's worth noting that other optimization criteria, such as ensuring fairness among users, could also be considered and are left for future work.}. The corresponding optimization problem can be formulated as
\begin{align}\label{Beamforming_Design_8}
    \mathcal{P}\mathrm{1}:&\max_{\mathbf{V}^{(1)},\mathbf{V}^{(2)},\cdots,\mathbf{V}^{(K)},
    \overline{\mathbf{\Theta}}^{(0)},\overline{\mathbf{\Theta}}^{(1)},\cdots,
    \overline{\mathbf{\Theta}}^{(L)}}\frac{1}{K}\sum_{k=1}^{K}\sum_{u=1}^{U}R_u^{(k)}\\
    \text{s.t.}&\quad \overline{\mathbf{\Theta}}^{(l)}
    \overline{\mathbf{\Theta}}^{(l)\mathrm{H}}=\mathbf{I}_N,\quad l=0,1,\cdots,L,\\
    &\quad \left\|\mathbf{v}_u^{(k)}\right\|^2=1,\quad k=1,2,\cdots,K,\quad u=1,2,\cdots,U,\\
    &\quad \rho_1^{(k)}+\rho_2^{(k)}+\cdots+\rho_U^{(k)}=\rho^{(k)},\quad k=1,2,\cdots,K.
\end{align}
Since $\mathcal{P}\mathrm{1}$ is a non-convex problem, we can decouple it into a pair of sub-problems and optimize them separately. Specifically, the SIM-based holographic beamformer is designed by our proposed layer-by-layer alternating optimization algorithm for maximizing the baseband channel gain, while the digital beamformer is optimized based on the MMSE precoding method for eliminating the inter-user interference and the SIM phase tuning error by leveraging their statistical information. This approach follows the hybrid beamforming architecture, where the RF-domain beamforming performed by SIM focuses on maximizing the equivalent channel power. This design philosophy aligns with the low-complexity requirements of phase-only control in the RF domain, leveraging the large aperture of SIM for efficiently enhancing the array gain~\cite{ni2015hybrid}. Meanwhile, the baseband digital beamforming employs computationally intensive methods, such as MMSE optimization, to suppress the inter-user interference and for enhancing the beamforming accuracy in the frequency domain. By dividing the optimization tasks between the RF and baseband domains, this layered strategy strikes a balance between computational efficiency and overall system performance, making it well-suited for practical implementations.

\subsection{SIM Phase Shift Optimization for Holographic Beamformer}
The objective function of maximizing the achievable sum-rate in $\mathcal{P}\mathrm{1}$ is tightly upper-bounded by that of maximizing the equivalent baseband eigen-channel gain of all UEs \cite{an2024adjustable}. Therefore, in this sub-problem of holographic beamforming, we aim for optimizing the SIM phase shifts $\overline{\mathbf{\Theta}}^{(0)},\overline{\mathbf{\Theta}}^{(1)},\cdots,
\overline{\mathbf{\Theta}}^{(L)}$ to maximize the sum of the equivalent baseband eigen-channel gain of all UEs over each subcarrier based on the sum-path-gain maximization (SPGM) criterion \cite{ning2020beamforming}, which can be formulated as
\begin{align}\label{Beamforming_Design_9}
    \mathcal{P}\mathrm{2a}:&\max_{\overline{\mathbf{\Theta}}^{(0)},\overline{\mathbf{\Theta}}^{(1)},
    \cdots,\overline{\mathbf{\Theta}}^{(L)}}
    \sum_{u=1}^{U}\left\|\overline{\mathbf{h}}_u^{(k)\mathrm{H}}\right\|^2\\
    \text{s.t.}&\quad \overline{\mathbf{\Theta}}^{(l)}
    \overline{\mathbf{\Theta}}^{(l)\mathrm{H}}=\mathbf{I}_N,
    \quad l=0,1,\cdots,L.
\end{align}
Note that all subcarriers share identical SIM phase shift designs $\overline{\mathbf{\Theta}}^{(0)},\overline{\mathbf{\Theta}}^{(1)},\cdots,
\overline{\mathbf{\Theta}}^{(L)}$. While the optimal SIM-based holographic beamformers can be designed for maximizing the spectral efficiency of a specific subcarrier, such a design may degrade the spectral efficiency of other subcarriers due to the frequency-selective nature of wideband channels. This issue becomes more severe as the frequency subcarrier spacing increases. To address this, we design our holographic beamformer by maximizing the spectral efficiency of the central carrier frequency. This choice is motivated by the need to simplify the optimization problem while maintaining acceptable performance across the band. Such a strategy is commonly employed in wideband systems as a practical starting point, as highlighted in~\cite{xu2024near}, where optimizing for the central carrier frequency serves as a foundation for addressing wideband effects. Although this approach primarily targets narrowband optimization, it lays the groundwork for future extensions to joint subcarrier-level optimization, which can comprehensively address dispersive wideband effects. Based on this design philosophy, problem $\mathcal{P}\mathrm{2a}$ can be reformulated as:
\begin{align}\label{Beamforming_Design_10}
    \mathcal{P}\mathrm{2b}:&\max_{\overline{\mathbf{\Theta}}^{(0)},\overline{\mathbf{\Theta}}^{(1)},
    \cdots,\overline{\mathbf{\Theta}}^{(L)}}
    \sum_{u=1}^{U}\left\|\overline{\mathbf{h}}_u^{(\mathrm{c})\mathrm{H}}\right\|^2\\
    \text{s.t.}&\quad \overline{\mathbf{\Theta}}^{(l)}
    \overline{\mathbf{\Theta}}^{(l)\mathrm{H}}=\mathbf{I}_N,
    \quad l=0,1,\cdots,L,
\end{align}
where $\overline{\mathbf{h}}_u^{(\mathrm{c})\mathrm{H}}$ represents the mean of the baseband equivalent channel at the central carrier frequency, given by
\begin{align}\label{Beamforming_Design_11}
    \notag\overline{\mathbf{h}}_u^{(\mathrm{c})\mathrm{H}}
    =&\xi^{L+1}\mathbf{g}_u^{(\mathrm{c})\mathrm{H}}\overline{\mathbf{\Theta}}^{(L)}
    \mathbf{F}_\mathrm{c}^{(L)}\overline{\mathbf{\Theta}}^{(L-1)}\mathbf{F}_\mathrm{c}^{(L-1)}
    \cdots\overline{\mathbf{\Theta}}^{(1)}\mathbf{F}_\mathrm{c}^{(1)}\\
    &\overline{\mathbf{\Theta}}^{(0)}\mathbf{F}_\mathrm{c}^{(0)}.
\end{align}
In (\ref{Beamforming_Design_11}), we have
\begin{align}\label{Beamforming_Design_12}
    \left[\mathbf{F}_\mathrm{c}^{(0)}\right]_{n,m}=\frac{1}{\sqrt{N}}\mathrm{e}^{-\jmath \frac{2\pi}{\lambda_\mathrm{c}}
    \left\|\mathbf{p}_{n}^{(0)}-\mathbf{q}_{m}\right\|},
\end{align}
\begin{align}\label{Beamforming_Design_13}
    \left[\mathbf{F}_\mathrm{c}^{(l)}\right]_{n_2,n_1}=\sqrt{\beta_{n_2,n_1}^{(l)}}
    \mathrm{e}^{-\jmath \frac{2\pi}{\lambda_\mathrm{c}}
    \left\|\mathbf{p}_{n_2}^{(l)}-\mathbf{p}_{n_1}^{(l-1)}\right\|},\quad l=1,2,\cdots L,
\end{align}
and
\begin{align}\label{Beamforming_Design_14}
    \notag\mathbf{g}_u^{(\mathrm{c})\mathrm{H}}=&\left[
    \sqrt{\zeta_{1}}\mathrm{e}^{-\jmath\frac{2\pi}{\lambda_\mathrm{c}}
    \left\|\mathbf{r}-\mathbf{p}_{1}^{(L)}\right\|},
    \sqrt{\zeta_{2}}\mathrm{e}^{-\jmath\frac{2\pi}{\lambda_\mathrm{c}}
    \left\|\mathbf{r}-\mathbf{p}_{2}^{(L)}\right\|},\right.\\
    &\left.\cdots,\sqrt{\zeta_{N}}\mathrm{e}^{-\jmath\frac{2\pi}{\lambda_\mathrm{c}}
    \left\|\mathbf{r}-\mathbf{p}_{N}^{(L)}\right\|}
    \right].
\end{align}

Since the sub-problem $\mathcal{P}\mathrm{2b}$ is still non-convex, we propose a layer-by-layer iterative optimization algorithm. At the beginning, we randomly set the initial SIM-based beamforming matrices $\overline{\mathbf{\Theta}}^{(0)},
\overline{\mathbf{\Theta}}^{(1)},\cdots,\overline{\mathbf{\Theta}}^{(L)}$. Based on this, we can get the equivalent deterministic baseband channels $\overline{\mathbf{h}}_1^{(\mathrm{c})\mathrm{H}},\overline{\mathbf{h}}_2^{(\mathrm{c})\mathrm{H}},
\cdots,\overline{\mathbf{h}}_U^{(\mathrm{c})\mathrm{H}}$. Then, the phase shift matrices of all SIM layers are optimized alternately. Specifically, firstly, we optimize the phase shift of $\overline{\mathbf{\Theta}}^{(0)}$ in layer-0 by fixing all the other $L$ layers of the SIM. Therefore, the baseband sum-path-gain can be represented as $\sum_{u=1}^{U}\|\ddot{\mathbf{h}}_{\mathrm{c},u}^{(0)\mathrm{H}}
\overline{\mathbf{\Theta}}^{(0)}\ddot{\mathbf{v}}_{\mathrm{c},u}^{(0)}\|^2$ in conjunction with the equivalent deterministic channel between the RF chains and
layer-0 being $\ddot{\mathbf{v}}_{\mathrm{c},u}^{(0)}
=\mathbf{F}_\mathrm{c}^{(0)}\overline{\mathbf{h}}_u^{(\mathrm{c})}$
and the equivalent determinate channel from layer-0 to the UE-$u$ being $\ddot{\mathbf{h}}_{\mathrm{c},u}^{(0)\mathrm{H}}
=\mathbf{g}_u^{(\mathrm{c})\mathrm{H}}\overline{\mathbf{\Theta}}^{(L)}
\mathbf{F}_\mathrm{c}^{(L)}\cdots\overline{\mathbf{\Theta}}^{(1)}\mathbf{F}_\mathrm{c}^{(1)}$. In layer-0, the optimization problem can be formulated as
\begin{align}\label{Beamforming_Design_holographic_4}
    \mathcal{P}\mathrm{3a}:&\max_{\overline{\mathbf{\Theta}}^{(0)}}
    \sum_{u=1}^{U}\left\|\ddot{\mathbf{h}}_{\mathrm{c},u}^{(0)\mathrm{H}}
    \overline{\mathbf{\Theta}}^{(0)}\ddot{\mathbf{v}}_{\mathrm{c},u}^{(0)}\right\|^2\\
    \text{s.t.}&\quad \overline{\mathbf{\Theta}}^{(0)}
    \overline{\mathbf{\Theta}}^{(0)\mathrm{H}}=\mathbf{I}_N.
\end{align}
By defining $\overline{\boldsymbol{\theta}}^{(0)}=[\mathrm{e}^{\jmath\overline{\theta}_1^{(0)}},
\mathrm{e}^{\jmath\overline{\theta}_2^{(0)}},\cdots,\mathrm{e}^{\jmath\overline{\theta}_N^{(0)}}
]^{\mathrm{T}}$ and $\mathbf{Z}^{(0)}=[\mathbf{z}_1^{(0)},\mathbf{z}_2^{(0)},\cdots,
\mathbf{z}_U^{(0)}]^{\mathrm{H}}$ with $\mathbf{z}_u^{(0)\mathrm{H}}=
\ddot{\mathbf{h}}_{\mathrm{c},u}^{(0)\mathrm{H}}\odot
\ddot{\mathbf{v}}_{\mathrm{c},u}^{(0)\mathrm{T}}$, problem $\mathcal{P}\mathrm{3a}$ becomes equivalent to
\begin{align}\label{Beamforming_Design_holographic_5}
    \mathcal{P}\mathrm{3b}:&\max_{\overline{\boldsymbol{\theta}}^{(0)}}\
    \overline{\boldsymbol{\theta}}^{(0)\mathrm{H}}\mathbf{Z}_{\mathrm{c}}^{(0)\mathrm{H}}
    \mathbf{Z}_{\mathrm{c}}^{(0)}\overline{\boldsymbol{\theta}}^{(0)}\\
    \text{s.t.}&\quad \left|\overline{\boldsymbol{\theta}}^{(0)}\right|=\mathbf{1}_N.
\end{align}
Searching for the solution in $\mathcal{P}\mathrm{3b}$ is an NP-hard problem. Although it can be solved by the classical semidefinite relaxation (SDR) method, it has high calculation complexity \cite{chi2017convex}. To circumvent this problem, we employ the low-complexity rank-one approximation method. Specifically, we can simply choose the principal eigenvector of the matrix $\mathbf{Z}_{\mathrm{c}}^{(0)\mathrm{H}}\mathbf{Z}_{\mathrm{c}}^{(0)}$ and then take its phase. Thus, the holographic beamforming matrix $\overline{\mathbf{\Theta}}^{(0)}$ can be optimized as $\overline{\mathbf{\Theta}}^{(0)}=\mathbf{Diag}\{{\mathrm{e}^{\jmath\angle\boldsymbol{\mu}^{(0)}}}\}$, where $\boldsymbol{\mu}^{(0)}$ is the eigen-vector of the matrix $\mathbf{Z}_{\mathrm{c}}^{(0)\mathrm{H}}\mathbf{Z}_{\mathrm{c}}^{(0)}$ associated with the maximal eigen-value. Afterwards, we optimize $\overline{\mathbf{\Theta}}^{(1)}$ by fixing all the other $L$ layers of the SIM with the baseband sum-path-gain represented as $\sum_{u=1}^{U}\|\ddot{\mathbf{h}}_{\mathrm{c},u}^{(1)\mathrm{H}}
\overline{\mathbf{\Theta}}^{(1)}\ddot{\mathbf{v}}_{\mathrm{c},u}^{(1)}\|^2$ along with the equivalent deterministic channel between the RF chains and layer-1 being $\ddot{\mathbf{v}}_{\mathrm{c},u}^{(1)}=\mathbf{F}_\mathrm{c}^{(1)}
\overline{\mathbf{\Theta}}^{(0)}\mathbf{F}_\mathrm{c}^{(0)}\overline{\mathbf{h}}_u^{(\mathrm{c})}$,
and the equivalent deterministic channel between layer-1 and UE-$u$ being $\ddot{\mathbf{h}}_{\mathrm{c},u}^{(1)\mathrm{H}}
=\mathbf{g}_u^{(\mathrm{c}){\mathrm{H}}}\overline{\mathbf{\Theta}}^{(L)}
\mathbf{F}_\mathrm{c}^{(L)}\cdots\overline{\mathbf{\Theta}}^{(2)}\mathbf{F}_\mathrm{c}^{(2)}$. The holographic beamforming matrix $\overline{\mathbf{\Theta}}^{(1)}$ can be similarly optimized according to $\overline{\mathbf{\Theta}}^{(1)}=
\mathbf{Diag}\{{\mathrm{e}^{\jmath\angle\boldsymbol{\mu}^{(1)}}}\}$, where $\boldsymbol{\mu}^{(1)}$ is the eigen-vector of the matrix $\mathbf{Z}_{\mathrm{c}}^{(1)\mathrm{H}}\mathbf{Z}_{\mathrm{c}}^{(1)}$ associated with the maximal eigen-value, where we have $\mathbf{Z}^{(1)}=[\mathbf{z}_1^{(1)},\mathbf{z}_2^{(1)},\cdots,
\mathbf{z}_U^{(1)}]^{\mathrm{H}}$ with $\mathbf{z}_u^{(1)\mathrm{H}}=
\ddot{\mathbf{h}}_{\mathrm{c},u}^{(1)\mathrm{H}}\odot
\ddot{\mathbf{v}}_{\mathrm{c},u}^{(1)\mathrm{T}}$. Similarly, the holographic beamforming matrices $\overline{\mathbf{\Theta}}^{(2)},\overline{\mathbf{\Theta}}^{(3)},\cdots,
\overline{\mathbf{\Theta}}^{(L)}$ can be optimized in turn. The details of the layer-by-layer iterative optimization of the hybrid beamformer are presented in Algorithm~\ref{algorithm_1}.

From lines 4 to 16 in Algorithm~\ref{algorithm_1}, the optimization goal is to maximize the channel gain of $\|\ddot{\mathbf{h}}_{\mathrm{c},u}^{(l)\mathrm{H}}
\overline{\mathbf{\Theta}}^{(l)}\ddot{\mathbf{v}}_{\mathrm{c},u}^{(l)}\|^2$. This is achieved by designing the SIM-based beamforming matrix $\overline{\mathbf{\Theta}}^{(l)}$ in layer-$l$ ($l=0,1,\cdots,L$). It is important to note that for each layer-$l$, the optimized SIM-based beamforming matrix yields a higher channel gain compared to the previous iteration. This is evident from line 19 of Algorithm~\ref{algorithm_1}, where the updated SIM-based beamforming matrix in layer-$l$ is designed for maximizing the channel gain by coherently combining the equivalent determinate channels $\ddot{\mathbf{h}}_{\mathrm{c},u}^{(l)}$ and $\ddot{\mathbf{v}}_{\mathrm{c},u}^{(l)}$. As a result, the channel gain $\|\ddot{\mathbf{h}}_{\mathrm{c},u}^{(l)\mathrm{H}}
\overline{\mathbf{\Theta}}^{(l)}\ddot{\mathbf{v}}_{\mathrm{c},u}^{(l)}\|^2$ increases monotonically with each iteration. Since the channel gain is inherently bounded, i.e., there exists an upper limit to the channel gain $\|\ddot{\mathbf{h}}_{\mathrm{c},u}^{(l)\mathrm{H}}
\overline{\mathbf{\Theta}}^{(l)}\ddot{\mathbf{v}}_{\mathrm{c},u}^{(l)}\|^2$, we may conclude that the algorithm converges.

\begin{algorithm}[!t]
\caption{Layer-by-layer iterative optimization algorithm for the SIM-based holographic beamformer.}
\label{algorithm_1}
\begin{algorithmic}[1]
\REQUIRE
    The channel links $\mathbf{F}_\mathrm{c}^{(0)},\mathbf{F}_\mathrm{c}^{(1)},\cdots,\mathbf{F}_\mathrm{c}^{(L)}$ and $\mathbf{g}_1^{(\mathrm{c})\mathrm{H}},\mathbf{g}_2^{(\mathrm{c})\mathrm{H}},\cdots,
    \mathbf{g}_U^{(\mathrm{c})\mathrm{H}}$, and the threshold $\epsilon$ used to determine the stopping criterion.
    \STATE
        Set the random initial SIM-based beamforming matrices $\overline{\mathbf{\Theta}}^{(0)}$, $\overline{\mathbf{\Theta}}^{(1)}$, $\cdots$, $\overline{\mathbf{\Theta}}^{(L)}$, satisfying $\overline{\mathbf{\Theta}}^{(0)}\overline{\mathbf{\Theta}}^{(0)\mathrm{H}}=\mathbf{I}_N$,
        $\overline{\mathbf{\Theta}}^{(1)}\overline{\mathbf{\Theta}}^{(1)\mathrm{H}}=\mathbf{I}_N$,
        $\cdots$, $\overline{\mathbf{\Theta}}^{(L)}\overline{\mathbf{\Theta}}^{(L)\mathrm{H}}=\mathbf{I}_N$.
    \REPEAT
    \FOR{$l=0$ to $L$}
    \FOR{$u=1$ to $U$}
        \STATE
            The equivalent determinate baseband channel of UE-$u$ is $\overline{\mathbf{h}}_u^{(\mathrm{c})\mathrm{H}}
                =\xi^{L+1}\mathbf{g}_u^{(\mathrm{c})\mathrm{H}}\overline{\mathbf{\Theta}}^{(L)}
                \mathbf{F}_\mathrm{c}^{(L)}\cdots\overline{\mathbf{\Theta}}^{(1)}
                \mathbf{F}_\mathrm{c}^{(1)}
                \overline{\mathbf{\Theta}}^{(0)}\mathbf{F}_\mathrm{c}^{(0)}$.
        \IF{$l=0$}
            \STATE
              The equivalent determinate channel from RF chains to layer-0 is  $\ddot{\mathbf{v}}_\mathrm{c,u}^{(0)}=
                \mathbf{F}_\mathrm{c}^{(0)}\overline{\mathbf{h}}_u^{(\mathrm{c})}$.
            \STATE
            The equivalent determinate channel from layer-0 to UE-$u$ is  $\ddot{\mathbf{h}}_\mathrm{c,u}^{(0)\mathrm{H}}
                =\xi^L\mathbf{g}_u^{(\mathrm{c})\mathrm{H}}\overline{\mathbf{\Theta}}^{(L)}
                \mathbf{F}_\mathrm{c}^{(L)}\cdots
                \overline{\mathbf{\Theta}}^{(1)}\mathbf{F}_\mathrm{c}^{(1)}$.
        \ELSIF{$l=1,2,\cdots,L-1$}
            \STATE
               The equivalent determinate channel from RF chains to layer-$l$ is $\ddot{\mathbf{v}}_\mathrm{c}^{(l)}=\xi^l\mathbf{F}_\mathrm{c}^{(l)}
                \overline{\mathbf{\Theta}}^{(l-1)}\mathbf{F}_\mathrm{c}^{(l-1)}\cdots
                \overline{\mathbf{\Theta}}^{(0)}\mathbf{F}_\mathrm{c}^{(0)}
                \overline{\mathbf{h}}_u^{(\mathrm{c})}$.
            \STATE
                The equivalent determinate channel from layer-$l$ to UE-$u$ is  $\ddot{\mathbf{h}}_\mathrm{c}^{(l)\mathrm{H}}=\xi^{L-l}
                \mathbf{g}_u^{(\mathrm{c})\mathrm{H}}
                \overline{\mathbf{\Theta}}^{(L)}\mathbf{F}_\mathrm{c}^{(L)}
                \cdots\overline{\mathbf{\Theta}}^{(l+1)}\mathbf{F}_\mathrm{c}^{(l+1)}$.
        \ELSIF{$l=L$}
            \STATE
                The equivalent determinate channel links from RF chains to layer-$L$ is   $\ddot{\mathbf{v}}_\mathrm{c}^{(L)}=\xi^L\mathbf{F}_\mathrm{c}^{(L)}
                \overline{\mathbf{\Theta}}^{(L-1)}\mathbf{F}_\mathrm{c}^{(L-1)}\cdots
                \overline{\mathbf{\Theta}}^{(0)}\mathbf{F}_\mathrm{c}^{(0)}
                \overline{\mathbf{h}}_u^{(\mathrm{c})}$.
            \STATE
                The equivalent determinate channel links from layer-$L$ to UE-$u$ is  $\ddot{\mathbf{h}}_\mathrm{c}^{(l)\mathrm{H}}=\mathbf{g}_u^{(\mathrm{c})\mathrm{H}}$.
        \ENDIF
    \ENDFOR
        \STATE
            Define the matrix $\mathbf{Z}^{(l)}=[\mathbf{z}_1^{(l)},\mathbf{z}_2^{(l)},\cdots,
             \mathbf{z}_U^{(l)}]^{\mathrm{H}}$ with $\mathbf{z}_u^{(l)\mathrm{H}}=\ddot{\mathbf{h}}_{\mathrm{c},u}^{(l)\mathrm{H}}
            \odot\ddot{\mathbf{v}}_{\mathrm{c},u}^{(l)\mathrm{T}}$.
        \STATE
            Calculate the eigen-vector of the matrix $\mathbf{Z}^{(l)\mathrm{H}}\mathbf{Z}^{(l)}$ associated with the maximal eigen-value, denoted as $\boldsymbol{\mu}^{(l)}$.
        \STATE
            The optimal SIM-based beamforming matrix for layer-$l$ is given by $\overline{\mathbf{\Theta}}^{(l)}=\mathbf{Diag}
            \{{\mathrm{e}^{\jmath\angle\boldsymbol{\mu}^{(l)}}}\}$.
    \ENDFOR
       \UNTIL{the absolute change in the objective function of $\sum_{u=1}^{U}\|\overline{\mathbf{h}}_u^{(\mathrm{c})\mathrm{H}}\|^2$ is smaller than the threshold $\epsilon$.}
\ENSURE
    The optimized SIM-based beamforming matrices $\overline{\mathbf{\Theta}}^{(0)},\overline{\mathbf{\Theta}}^{(1)},\cdots,
    \overline{\mathbf{\Theta}}^{(L)}$.
\end{algorithmic}
\end{algorithm}

\subsection{Precoding Vector Optimization for Digital Beamformers}
Once the SIM-based holographic beamformers $\overline{\mathbf{\Theta}}^{(0)},\overline{\mathbf{\Theta}}^{(1)},\cdots,
\overline{\mathbf{\Theta}}^{(L)}$ are obtained by leveraging the layer-by-layer iterative optimization algorithm, we can get the mean of the baseband equivalent channel $\mathbf{h}_u^{(k)\mathrm{H}}$ in (\ref{Beamforming_Design_5}) and the covariance matrix of the random channel uncertainty $\widetilde{\mathbf{h}}_u^{(k)\mathrm{H}}$ in (\ref{Beamforming_Design_6}). Problem $\mathcal{P}\mathrm{1}$ can be further reformulated as
\begin{align}\label{Beamforming_Design_9}
    \notag\mathcal{P}\mathrm{4}:&\max_{\mathbf{V}^{(1)},\mathbf{V}^{(2)},\cdots,\mathbf{V}^{(K)}}
    \frac{1}{K}\sum_{k=1}^{K}\sum_{u=1}^{U}\\
    &\qquad\log_2\left(1+\frac{\rho_u^{(k)}\left\|\overline{\mathbf{h}}_u^{(k)\mathrm{H}}
    \mathbf{S}^{(k)}\mathbf{v}_u^{(k)}\right\|^2}{\mathbf{v}_u^{(k)\mathrm{H}}
    \mathbf{Q}_u^{(k)}\mathbf{v}_u^{(k)}+\sigma_w^2}\right)\\
    \text{s.t.}&\quad \left\|\mathbf{v}_u^{(k)}\right\|^2=1,\quad k=1,2,\cdots,K,\quad u=1,2,\cdots,U,\\
    &\quad \rho_1^{(k)}+\rho_2^{(k)}+\cdots+\rho_U^{(k)}=\rho^{(k)},\quad k=1,2,\cdots,K.
\end{align}
By leveraging the generalized Rayleigh quotient for problem $\mathcal{P}\mathrm{4}$, the optimal digital beamformers over the $k$th subcarrier $\mathbf{v}_1^{(k)},\mathbf{v}_2^{(k)},\cdots,\mathbf{v}_U^{(k)}$ considering the inter-user interference and the effect of the SIM phase tuning errors can be designed by relying on the MMSE criterion as follows:
\begin{align}\label{Beamforming_Design_9_2}
    \mathbf{v}_u^{(k)}=\frac{\left(\mathbf{Q}_u^{(k)}+\sigma_w^2\mathbf{I}_M\right)^{-1}
    \mathbf{S}^{(k)\mathrm{H}}
    \overline{\mathbf{h}}_u^{(k)}}{\left\|\left(\mathbf{Q}_u^{(k)}+\sigma_w^2\mathbf{I}_M\right)^{-1}
    \mathbf{S}^{(k)\mathrm{H}}\overline{\mathbf{h}}_u^{(k)}\right\|}.
\end{align}
Based on this, the spectral efficiency of UE-$u$ over the $k$th subcarrier in (\ref{Beamforming_Design_7}) can be formulated as
\begin{align}\label{Power_Allocation_1}
    \notag R_u^{(k)}=&\log_2\left(1+\right.\\
    &\left. p_u\rho^{(k)}\overline{\mathbf{h}}_u^{(k)\mathrm{H}}
    \mathbf{S}^{(k)}\left(\mathbf{Q}_u^{(k)}+\sigma_w^2\mathbf{I}_M\right)^{-1}
    \mathbf{S}^{(k)\mathrm{H}}\overline{\mathbf{h}}_u^{(k)}\right),
\end{align}
where $p_u$ is the power sharing ratio of UE-$u$ satisfying $\sum_{u=1}^Up_u=1$. Although the conventional waterfilling algorithm can be employed for maximizing the achievable sum-rate, it is sub-optimum in the presence of inter-user interference and signal distortion resulting from the phase tuning error. This is due to the fact that the interference matrix $\mathbf{Q}_u^{(k)}$ also includes the power allocation ratio $p_1,p_2,\cdots,p_U$. Here, we can employ the iterative waterfilling power allocation algorithm. Specifically, given an arbitrary initial power allocation ratio, denoted as $p_1^{(0)},p_2^{(0)},\cdots,p_U^{(0)}$, the interference matrices $\mathbf{Q}_1^{(k)},\mathbf{Q}_2^{(k)},\cdots,\mathbf{Q}_U^{(k)}$ perceived by all users can be calculated. In the $t$th iteration, we treat the interference matrices $\mathbf{Q}_1^{(k)},\mathbf{Q}_2^{(k)},\cdots,\mathbf{Q}_U^{(k)}$ depending on the power sharing ratio in the $(t-1)$st iteration as noise, and then the power sharing ratio in the $t$th iteration, denoted as $p_1^{(t)},p_2^{(t)},\cdots,p_U^{(t)}$ can be calculated by the conventional waterfilling algorithm as follows:
\begin{align}\label{Power_Allocation_2}
    \notag p_u^{(t)}=&\left[\varrho^{(t-1)}-\right.\\
    &\left.\frac{1}{\rho^{(k)}\overline{\mathbf{h}}_u^{(k)\mathrm{H}}
    \mathbf{S}^{(k)}\left(\mathbf{Q}_u^{(k,t-1)}+\sigma_w^2\mathbf{I}_M\right)^{-1}
    \mathbf{S}^{(k)\mathrm{H}}\overline{\mathbf{h}}_u^{(k)}}\right]^{+},
\end{align}
where the interference matrix $\mathbf{Q}_u^{(k,t-1)}$ is calculated by the power allocation ratio in the $(t-1)$st iteration as
\begin{align}\label{Power_Allocation_3}
    \notag&\mathbf{Q}_u^{(k,t-1)}=p_u^{(t-1)}\rho^{(k)}\mathbf{S}^{(k)\mathrm{H}}
    \mathbf{C}_{\widetilde{\mathbf{h}}_u^{(k)}\widetilde{\mathbf{h}}_u^{(k)\mathrm{H}}}\mathbf{S}^{(k)}\\
    &+\sum_{u'=1,u'\neq u}^{U}p_{u'}^{(t-1)}\rho^{(k)}
    \mathbf{S}^{(k)\mathrm{H}}\left(\overline{\mathbf{h}}_{u'}^{(k)}\overline{\mathbf{h}}_{u'}^{(k)\mathrm{H}}
    +\mathbf{C}_{\widetilde{\mathbf{h}}_{u'}^{(k)}\widetilde{\mathbf{h}}_{u'}^{(k)\mathrm{H}}}\right)
    \mathbf{S}^{(k)},
\end{align}
and $\varrho^{(t-1)}$ is given by
\begin{align}\label{Power_Allocation_4}
    \notag&\varrho^{(t-1)}=\frac{1}{U}\left[1+\right.\\
    &\left.\sum_{u=1}^{U}\frac{1}{\rho^{(k)}\overline{\mathbf{h}}_u^{(k)
    \mathrm{H}}\mathbf{S}^{(k)}(\mathbf{Q}_u^{(k,t-1)}+\sigma_w^2\mathbf{I}_M)^{-1}
    \mathbf{S}^{(k)\mathrm{H}}\overline{\mathbf{h}}_u^{(k)}}\right].
\end{align}

\section{Performance Analysis}\label{Performance_Analysis}
In this section, we characterize the spectral efficiency performance of our proposed hybrid beamforming architecture relying on the SIM-based transceiver design.

Upon optimizing the digital beamformer and the holographic beamformer, the average spectral efficiency of UE-$u$, denoted as $R_u$, can be formulated as
\begin{align}\label{Performance_Analysis_1}
    \notag &R_u=\frac{1}{K}\sum_{k=1}^K\log_2\left(1+\rho_u^{(k)}\mathbf{g}_u^{(k)\mathrm{H}}
        \overline{\mathbf{A}}^{(k)}\mathbf{S}^{(k)}\right.\\
        \notag&\left.\left(
        \rho_u^{(k)}\mathbf{S}^{(k)\mathrm{H}}\mathbf{C}_{\widetilde{\mathbf{h}}_u^{(k)}
        \widetilde{\mathbf{h}}_u^{(k)}}\mathbf{S}^{(k)}
        +\sum_{u'=1,u'\neq u}^{U}\rho_{u'}^{(k)}
        \mathbf{S}^{(k)\mathrm{H}}\left(\overline{\mathbf{h}}_{u'}^{(k)}
        \overline{\mathbf{h}}_{u'}^{(k)\mathrm{H}}\right.\right.\right.\\
        &\left.\left.\left.+\mathbf{C}_{\widetilde{\mathbf{h}}_{u'}^{(k)}
        \widetilde{\mathbf{h}}_{u'}^{(k)}}\right)\mathbf{S}^{(k)}+\sigma_w^2\mathbf{I}_M\right)^{-1}
        \mathbf{S}^{(k)\mathrm{H}}\overline{\mathbf{A}}^{(k)\mathrm{H}}\mathbf{g}_u^{(k)}\right),
\end{align}
with $\mathbf{C}_{\widetilde{\mathbf{h}}_u^{(k)}\widetilde{\mathbf{h}}_u^{(k)}}$ given in (\ref{Beamforming_Design_6}).

In the high transmit power region, the average spectral efficiency of the SIM-aided system can be derived as follows.

\begin{theorem}\label{Theorem_2}
When the transmit power obeys $\rho^{(k)}\rightarrow\infty$ ($k=1,2,\cdots,K$), the achievable sum-rate, denoted as $R^{(p\rightarrow\infty)}$, is given by
\begin{align}\label{Performance_Analysis_2}
    \notag&R^{(p\rightarrow\infty)}=\frac{1}{K}\sum_{k=1}^K\sum_{u=1}^U\log_2\left(1+\xi^{2(L+1)}
    \mathbf{g}_u^{(k)\mathrm{H}}\overline{\mathbf{A}}^{(k)}\mathbf{S}^{(k)}\right.\\
    \notag&\left.\left(\mathbf{S}^{(k)\mathrm{H}}\mathbf{C}_{\widetilde{\mathbf{h}}_u^{(k)}
    \widetilde{\mathbf{h}}_u^{(k)}}\mathbf{S}^{(k)}+\sum_{u'=1,u'\neq u}^{U}\frac{p_{u'}}{p_u}
    \mathbf{S}^{(k)\mathrm{H}}\left(\overline{\mathbf{h}}_{u'}^{(k)}
    \overline{\mathbf{h}}_{u'}^{(k)\mathrm{H}}\right.\right.\right.\\
    &\left.\left.\left.+\mathbf{C}_{\widetilde{\mathbf{h}}_{u'}^{(k)}
    \widetilde{\mathbf{h}}_{u'}^{(k)}}\right)\mathbf{S}^{(k)}\right)^{-1}
    \mathbf{S}^{(k)\mathrm{H}}\overline{\mathbf{A}}^{(k)\mathrm{H}}\mathbf{g}_u^{(k)}\right).
\end{align}
\end{theorem}
\begin{IEEEproof}
    It can be obtained by setting $\rho^{(k)}\rightarrow\infty$ in (\ref{Performance_Analysis_1}).
\end{IEEEproof}

\begin{remark}\label{Remark_1}
Theorem \ref{Theorem_2} indicates that the average spectral efficiency is limited by the hardware quality of the SIM elements. Hence, the spectral efficiency saturates in the high transmit power region, when the hardware quality is non-ideal.
\end{remark}

Next, we investigate the effect of the inter-layer distance $d_0,d_1,\cdots,d_L$ on the average spectral efficiency.

\begin{theorem}\label{Theorem_3}
When the SIM layers are extremely close, i.e $d_l\rightarrow 0$ for $l=0,1,\cdots,L$, the achievable sum-rate, denoted as $R^{(d\rightarrow 0)}$, can be expressed as
\begin{align}\label{Performance_Analysis_3}
    \notag &R^{(d\rightarrow 0)}=\frac{1}{K}\sum_{k=1}^K\sum_{u=1}^U\log_2
    \left(1+\xi^{2(L+1)}\rho_u^{(k)}\mathbf{g}_u^{(k)\mathrm{H}}
    \mathbf{\Xi}_k\mathbf{F}_k^{(0)}\mathbf{S}^{(k)}\right.\\
    \notag&\left.\left(\left(1-\xi^{2(L+1)}\right)\sum_{u'=1}^{U}\rho_{u'}^{(k)}
    \mathbf{S}^{(k)\mathrm{H}}\mathbf{F}_k^{(0)\mathrm{H}}
    \left(\left(\mathbf{g}_{u'}^{(k)}\mathbf{g}_{u'}^{(k)\mathrm{H}}\right)
    \odot\mathbf{I}_N\right)\right.\right.\\
    \notag&\left.\left.\mathbf{F}_k^{(0)}\mathbf{S}^{(k)}
    +\xi^{2(L+1)}\sum_{u'=1,u'\neq u}^{U}\rho_{u'}^{(k)}
    \mathbf{S}^{(k)\mathrm{H}}\mathbf{F}_k^{(0)\mathrm{H}}
    \mathbf{\mathbf{\Xi}}_k^{\mathrm{H}}\mathbf{g}_{u'}^{(k)}\right.\right.\\
    &\left.\left.\mathbf{g}_{u'}\mathbf{\mathbf{\Xi}}_k\mathbf{F}_k^{(0)}\mathbf{S}^{(k)}
    +\sigma_w^2\mathbf{I}_M\right)^{-1}\mathbf{S}^{(k)\mathrm{H}}\mathbf{F}_k^{(0)\mathrm{H}}
    \mathbf{\mathbf{\Xi}}_k^{\mathrm{H}}\mathbf{g}_u^{(k)}\right),
\end{align}
where we have:
\begin{align}\label{Performance_Analysis_4}
    \notag\mathbf{\Xi}_k=&\mathbf{Diag}
    \left\{\mathrm{e}^{-\jmath \left(\overline{\theta}_1^{(0)}
    +\sum_{l=1}^{L}\left(\overline{\theta}_1^{(l)}-\frac{2\pi}{\lambda_k}\left\|\mathbf{p}_{1}^{(l)}
    -\mathbf{p}_{1}^{(l-1)}\right\|\right)\right)},\right.\\
    \notag&\left.\mathrm{e}^{-\jmath \left(\overline{\theta}_2^{(0)}
    +\sum_{l=1}^{L}\left(\overline{\theta}_2^{(l)}-\frac{2\pi}{\lambda_k}\left\|\mathbf{p}_{2}^{(l)}
    -\mathbf{p}_{2}^{(l-1)}\right\|\right)\right)},\cdots,\right.\\
    &\left.\mathrm{e}^{-\jmath \left(\overline{\theta}_N^{(0)}
    +\sum_{l=1}^{L}\left(\overline{\theta}_N^{(l)}-\frac{2\pi}{\lambda_k}\left\|\mathbf{p}_{N}^{(l)}
    -\mathbf{p}_{N}^{(l-1)}\right\|\right)\right)}\right\}.
\end{align}
\end{theorem}
\begin{IEEEproof}
    See Appendix \ref{Appendix_B}.
\end{IEEEproof}

\begin{remark}\label{Remark_2}
Theorem \ref{Theorem_3} shows that when the inter-layer distance obeys $d_l\rightarrow 0$, the optimal beamforming matrices $\overline{\mathbf{\Theta}}^{(L)}\cdots\overline{\mathbf{\Theta}}^{(1)}
\overline{\mathbf{\Theta}}^{(0)}$ are not unique, as long as they satisfy that $\sum_{l=0}^{L}\theta_n^{(l)}=\frac{2\pi}{\lambda_\mathrm{c}}
\sum_{l=1}^{L}\|\mathbf{p}_{n}^{(l)}-\mathbf{p}_{n}^{(l-1)}\|$ for $n=1,2,\cdots,N$ based on (\ref{Performance_Analysis_4}). Furthermore, according to (\ref{Performance_Analysis_3}), we can observe that the average spectral efficiency degrades with the increase of the number of SIM layers when $d_l\rightarrow 0$. This is due to the fact that the fully-connected structure of reconfigurable elements between adjacent layers will be destroyed, when the SIM inter-layer distance obeys $d_l\rightarrow 0$. This leads to a reduction in the degree of optimization freedom and cannot be compensated by increasing the number of SIM layers. On a similar note, increasing the number of SIM layers aggravates the effect of phase tuning error on the spectral efficiency performance.
\end{remark}

\section{Numerical and Simulation Results}\label{Numerical_and_Simulation_Results}
\begin{figure}[!t]
    \centering
    \includegraphics[width=3in]{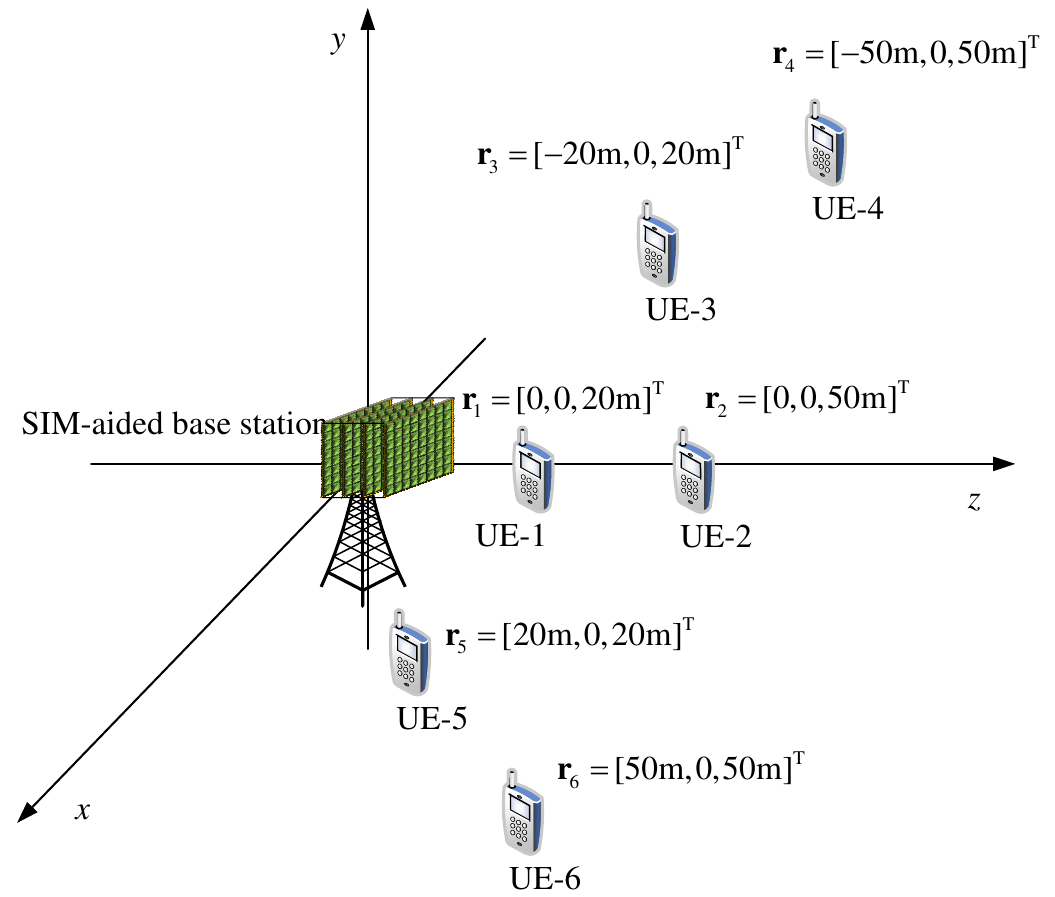}
    \caption{System setup of the simulation scenario.}\label{Fig_system_setup}
\end{figure}

In this section, the average spectral efficiency of the SIM-based transceiver designed for wideband systems is quantified. Unless otherwise specified, as in \cite{deng2022reconfigurable_twc}, \cite{cheng2024achievable} the simulation parameters are: the carrier frequency is $f_c=10$ GHz, the bandwidth is $B=600$ MHz, the number of subcarriers is $K=64$, the number of RF chains is $M=4\times4$ and $M=8\times8$ for the single-user case and the multi-user case, respectively. Furthermore, the number of SIM elements in each layer is $N=256\times256$, the total transmit power is $\rho=0\mathrm{dBm}$, the noise density is $\sigma_w^2=-104\mathrm{dBm/Hz}$, the phase tuning error variance is $\sigma_\mathrm{p}^2=0$ in the absence of phase tuning errors, and the size of each reconfigurable element is $\delta_x=\delta_y=\frac{\lambda_\mathrm{c}}{4}$. We assume that the inter-layer distance is $d_0=d_1=\cdots=d_L=5\lambda_\mathrm{c}$, and the feeds are evenly distributed on the input layer of the SIM. The number of iterations of the holographic beamforming algorithm is $\tau=4$. We assume that the mutual coupling effects between RF chain ports are neglected, i.e. $Z^{(k)}=Z_A\mathbf{I}_M$. This assumption simplifies the analysis and allows for a clearer evaluation of the fundamental performance trends of the proposed system. The system setup is shown in Fig. \ref{Fig_system_setup}. Specifically, the SIM-aided BS is located at the origin of $[0,0,0]^\mathrm{T}$ to support three pairs of UEs. In the first pair, UE-1 and UE-2 are at the same angle with the coordinate of $\mathbf{r}_1=[0,0,20\mathrm{m}]^\mathrm{T}$ and $\mathbf{r}_2=[0,0,50\mathrm{m}]^\mathrm{T}$, respectively. Similarly, in the second pair of users, UE-3 and UE-4 are at the same angle with the coordinate of $\mathbf{r}_3=[-20\mathrm{m},0,20\mathrm{m}]^\mathrm{T}$ and $\mathbf{r}_4=[-50\mathrm{m},0,50\mathrm{m}]^\mathrm{T}$, respectively. In the third pair of users, UE-5 and UE-6 are at the same angle with the coordinate of $\mathbf{r}_5=[20\mathrm{m},0,20\mathrm{m}]^\mathrm{T}$ and $\mathbf{r}_6=[50\mathrm{m},0,50\mathrm{m}]^\mathrm{T}$, respectively.

\begin{figure}[!t]
    \centering
    \includegraphics[width=2.8in]{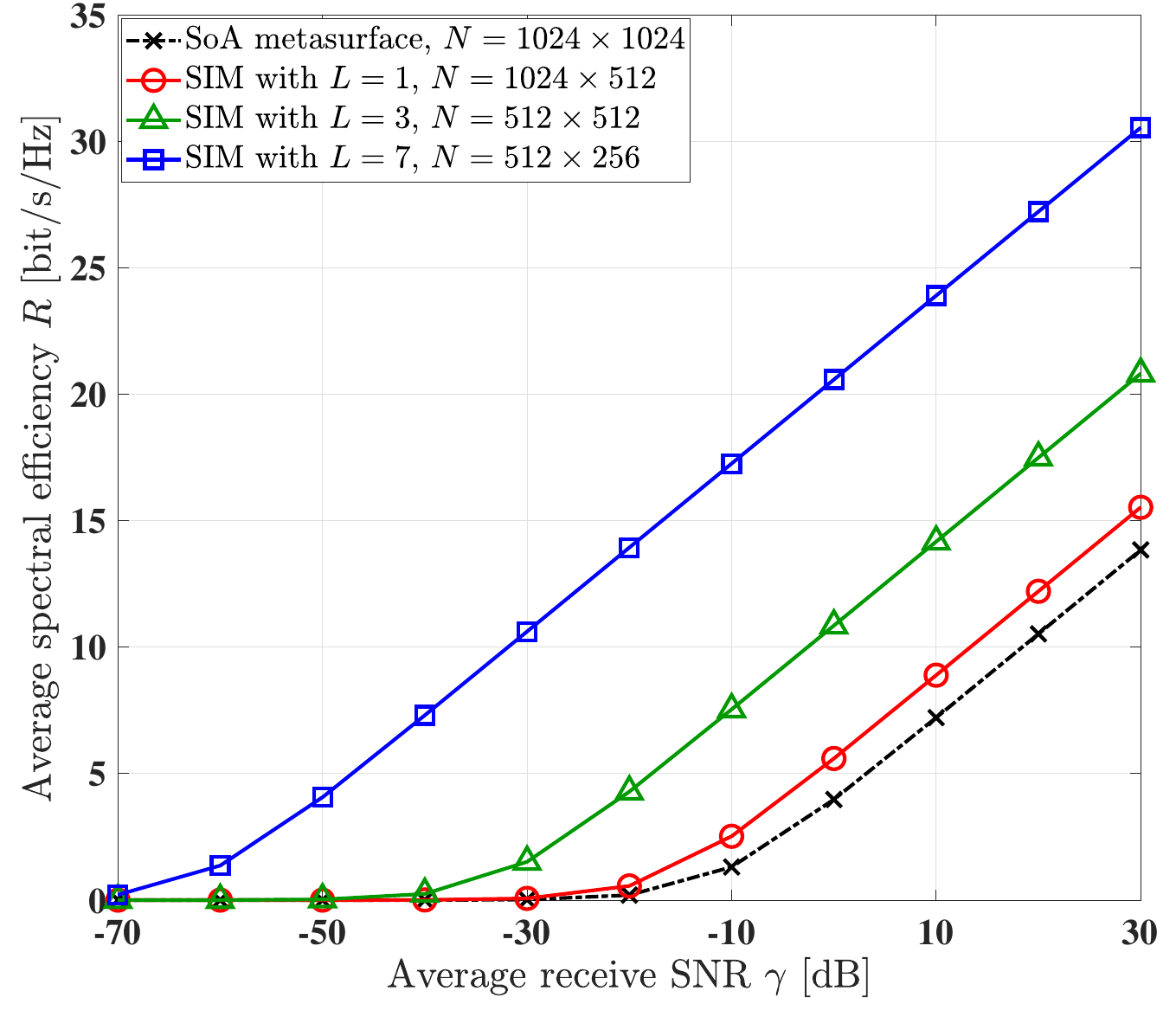}
    \caption{The average spectral efficiency $R$ versus the average receive SNR $\gamma$ in our proposed SIM-aided transceiver design and the SoA metasurface.}\label{Simu_Fig_perfect_1}
\end{figure}

\begin{figure}[!t]
    \centering
    \includegraphics[width=2.8in]{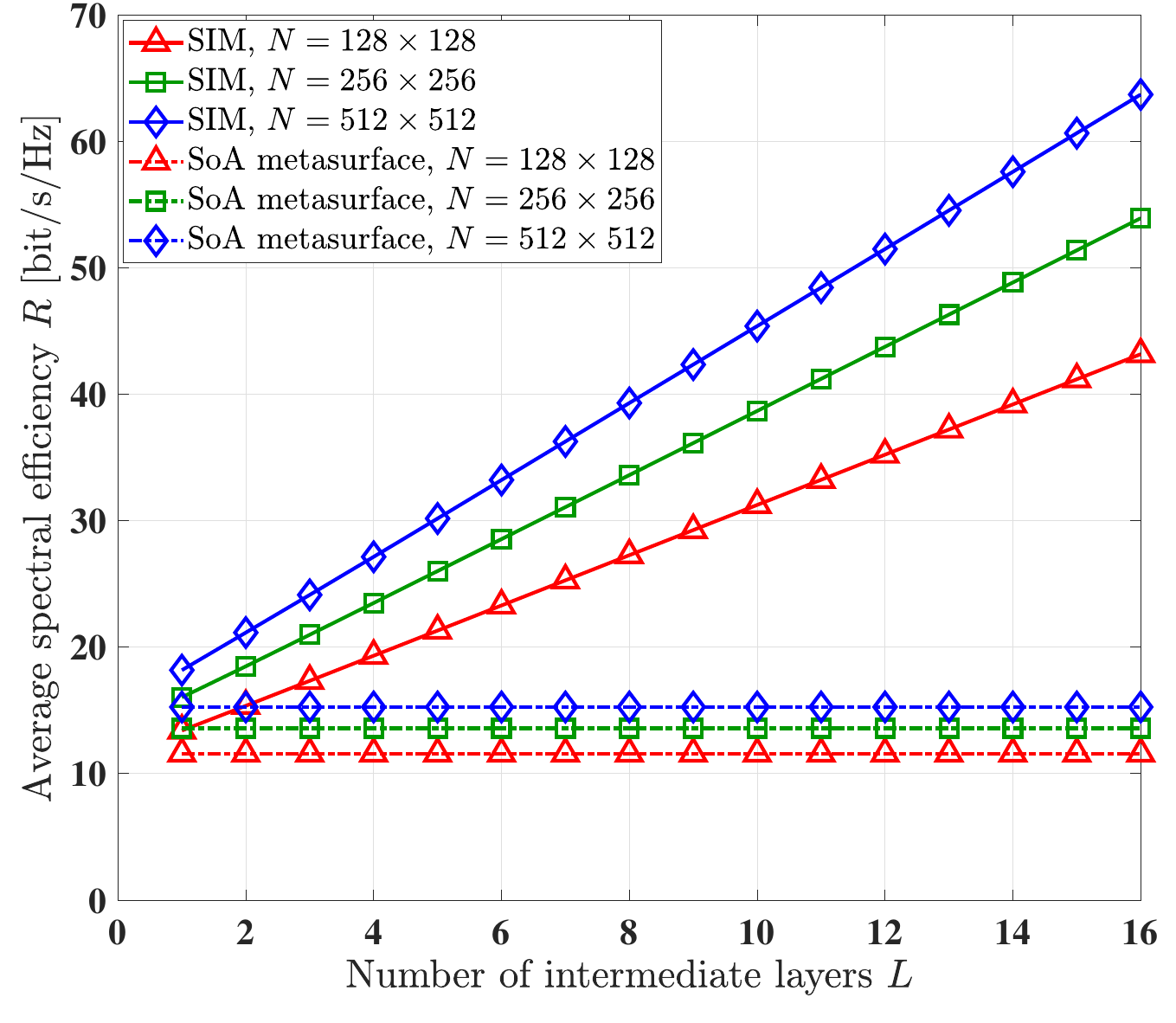}
    \caption{The average spectral efficiency $R$ versus the number of intermediate layers $L$ in the SIM, with different number of elements $N$ in each layer.}\label{Simu_Fig_perfect_2}
\end{figure}

\begin{figure}[!t]
    \centering
    \includegraphics[width=2.8in]{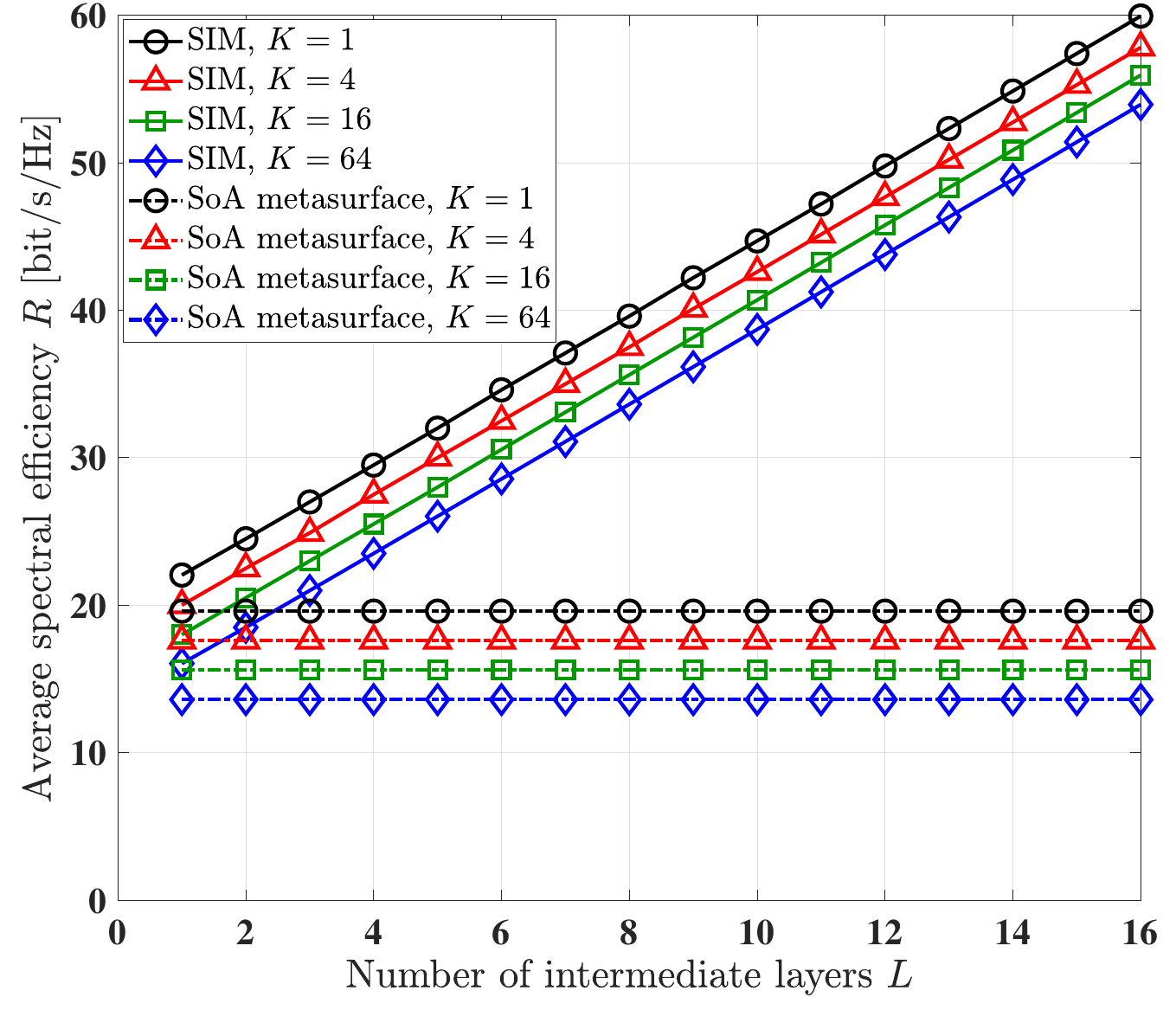}
    \caption{Comparison of the average spectral efficiency $R$ versus the number of intermediate layers $L$ in the SIM, with different number of subcarriers $K$.}\label{Simu_Fig_perfect_3}
\end{figure}

\begin{figure}[!t]
    \centering
    \includegraphics[width=2.8in]{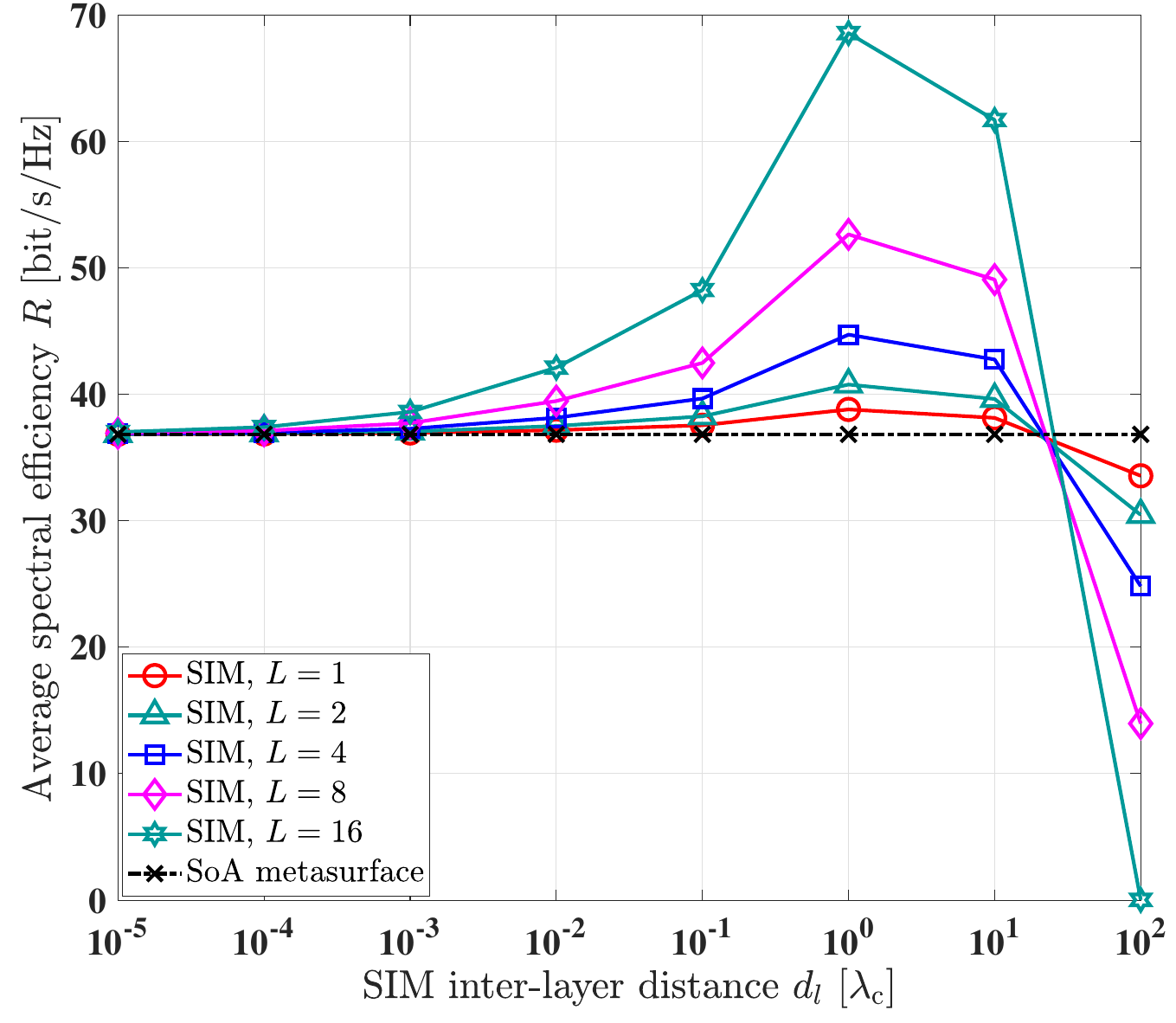}
    \caption{Comparison of the average spectral efficiency $R$ versus the SIM inter-layer distance $d_l$.}\label{Simu_Fig_perfect_4}
\end{figure}

\subsection{Performance Comparison of SIM-aided Transceiver Design and SoA Single-layer Metasurface}
As shown in Fig. \ref{Fig_system_setup}, in the single-user case we assume that the SIM-aided BS supports UE-2 in the wideband system.

Firstly, as shown in Fig. \ref{Simu_Fig_perfect_1}, we compare the average spectral efficiency $R$ versus the average receive SNR $\gamma$ in our proposed SIM-aided transceiver design and the SoA single-layer metasurface, i.e., $L=0$. The average receive SNR is defined as $\gamma=\frac{\rho\varrho}{\sigma_w^2}$, where $\varrho$ is the average path loss between the BS and the UE, which can be described as $\varrho=\mathrm{C}_0\|\mathbf{r}\|^{-2}$. Furthermore, $\mathrm{C}_0=-30\mathrm{dB}$ denotes the path loss at the
reference distance of 1 meter~\cite{wu2021intelligent}. Note that to ensure each scheme has the same total number of reconfigurable radiation elements, the number of elements in the SoA metasurface is $N=1024\times1024$. Furthermore, that in each layer of the SIM is set as $N=1024\times512$, $N=512\times512$ and $N=512\times256$ when the number of intermediate layers is $L=1$, $L=3$ and $L=7$, respectively. Fig. \ref{Simu_Fig_perfect_1} shows that the SIM-aided transceiver outperforms the SoA metasurface. Moreover, the achievable rate of the SIM-aided transceiver can be improved upon increasing the number of intermediate layers. This is due to the fact that more sophisticated connections may be realized for the holographic beamformer upon increasing the number of SIM layers.

Next, Fig. \ref{Simu_Fig_perfect_2} portrays the achievable rate $R$ versus the number of intermediate layers $L$ in the SIM architecture, parameterized by the different number of elements $N$ in each layer. Fig. \ref{Simu_Fig_perfect_2} shows that employing more reconfigurable radiation elements in each layer of the SIM-aided transceiver can promise higher achievable rate, albeit at the cost of increased hardware complexity and calculation complexity.

Furthermore, Fig. \ref{Simu_Fig_perfect_3} presents the average spectral efficiency $R$ versus the number of intermediate layers $L$ in the SIM architecture, with different number of subcarriers $K$. Observe that there is a performance degradation upon increasing the number of subcarriers. This can be explained as follows: the holographic beamformer is designed based on the criterion of maximizing the spectral efficiency at the central carrier frequency. The highest data rate can be achieved when the number of subcarriers is $K=1$, since all the transmit power is allocated to the subchannel at the central carrier frequency. By contrast, more transmit power is allocated to the subcarriers away from the central carrier frequency upon increasing the number of subcarriers.

Fig.~\ref{Simu_Fig_perfect_4} compares the average spectral efficiency $\overline{R}$ with respect to the SIM inter-layer distance. It shows that as the inter-layer distance approaches 0, the average spectral efficiency of the SIM architecture converges to that of the SoA metasurface, thereby validating Theorem~\ref{Theorem_3}. This occurs because, as the inter-layer distance obeys $d_l\rightarrow0$, the fully-connected structure of the SIM is destroyed. Specifically, as the distance between the SIM layers decreases, the signal from a reconfigurable element in one layer primarily propagates to the corresponding element in the adjacent layer, instead of being distributed across all elements in the next layer. This results in a significant degradation of the intended spatially distributed signal processing. By contrast, when the SIM inter-layer distance is relatively large, e.g. $d_l=100\lambda_\mathrm{c}$, increasing the number of SIM layers causes a decrease in the average spectral efficiency due to signal attenuation as it propagates through each layer, with each additional layer contributing to further degradation. Therefore, the results in Fig.~\ref{Simu_Fig_perfect_4} suggest the existence of an optimal inter-layer distance that maximizes the average spectral efficiency of the SIM architecture.

\begin{figure}[!t]
    \centering
    \subfloat[For different number of intermediate layers in the SIM architecture.]
    {\begin{minipage}{1\linewidth}
        \centering
        \includegraphics[width=2.8in]{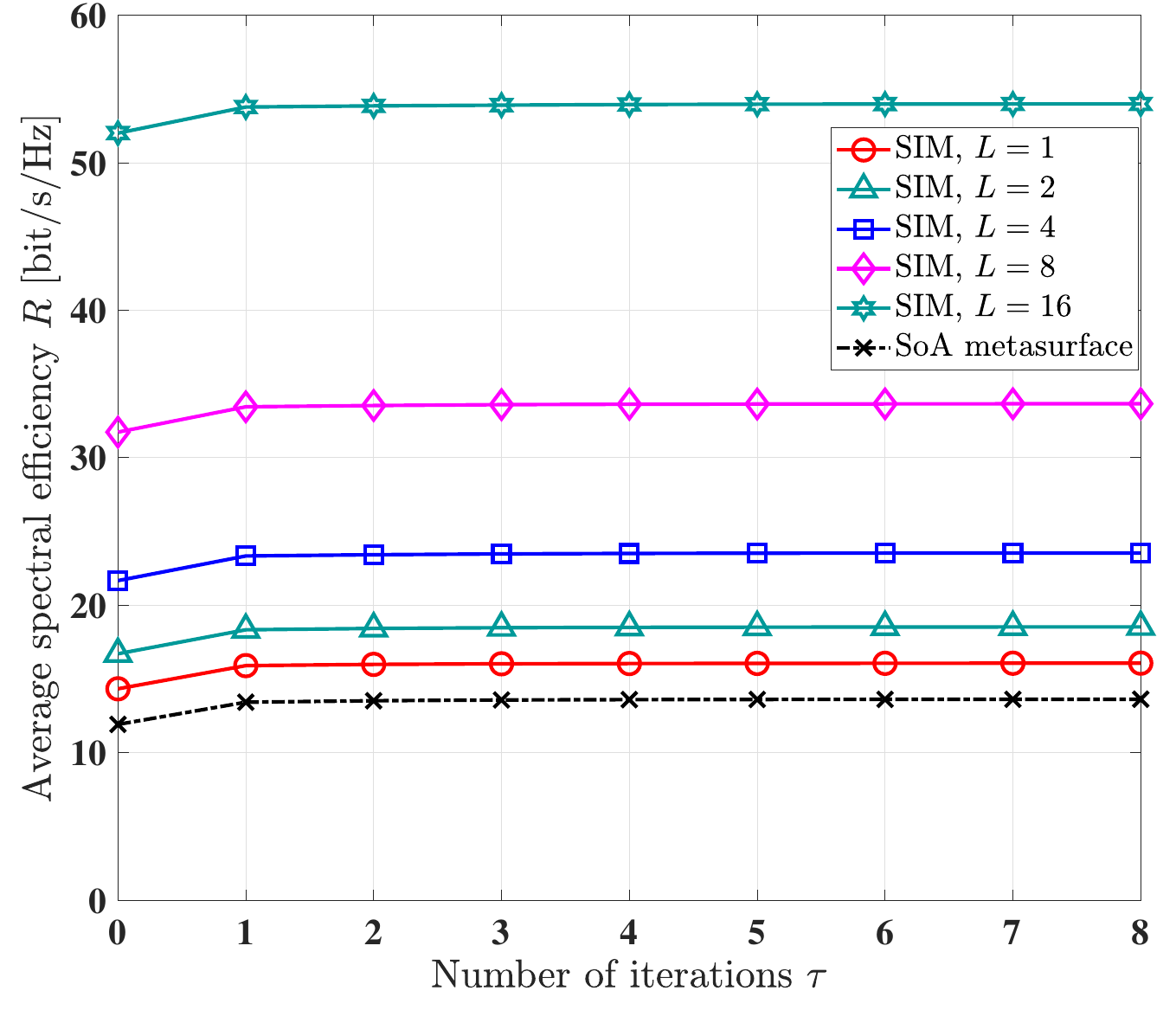}
    \end{minipage}}\\
    \subfloat[For different number of reconfigurable radiation elements in each layer.]
    {\begin{minipage}{1\linewidth}
        \centering
        \includegraphics[width=2.8in]{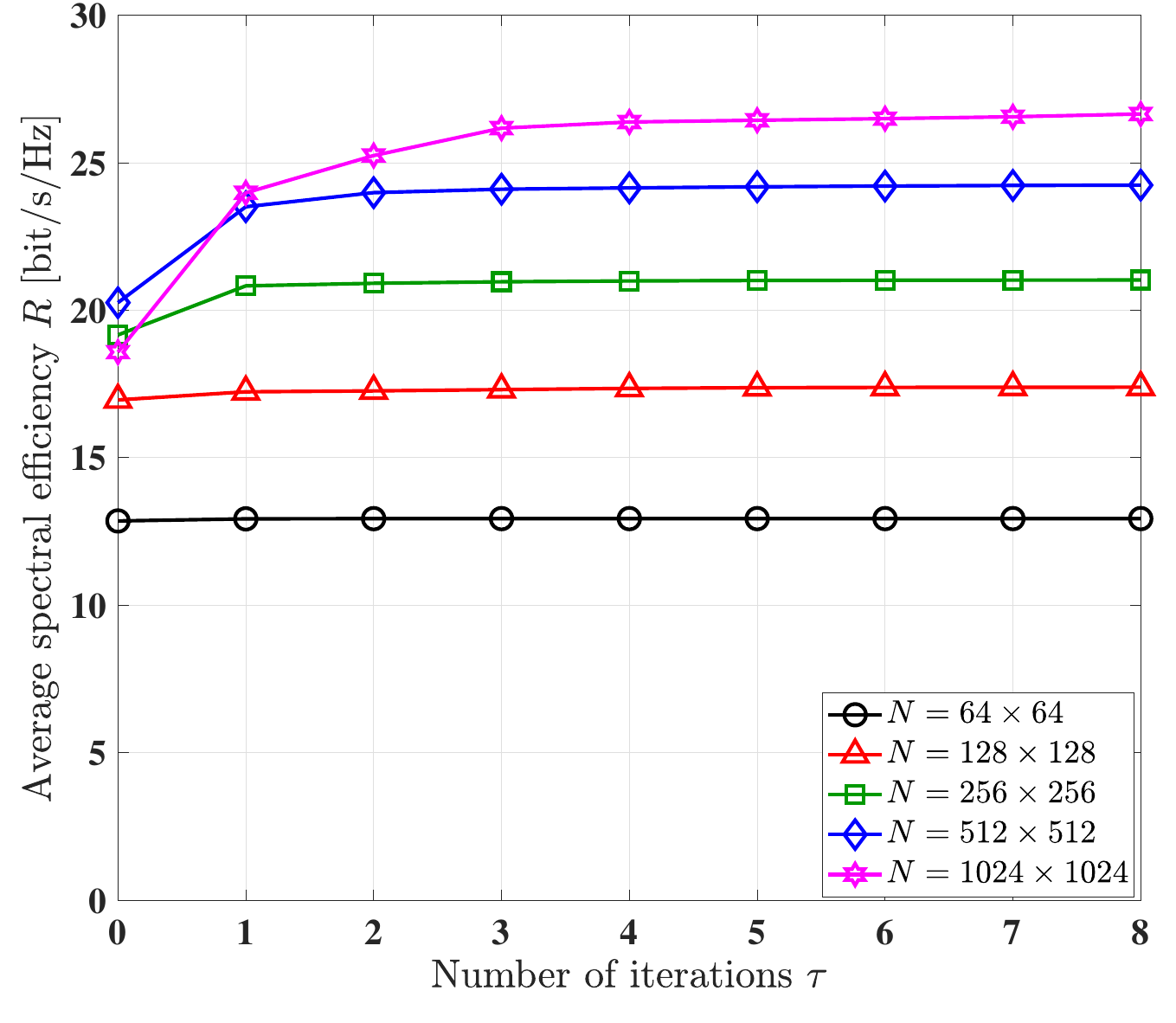}
    \end{minipage}}
    \caption{Comparison of the average spectral efficiency $R$ versus the number of iterations $\tau$ of the proposed layer-by-layer iterative optimization algorithm.}\label{Simu_Fig_perfect_56}
\end{figure}

In Fig. \ref{Simu_Fig_perfect_56} (a), the average spectral efficiency $R$ versus the number of iterations $\tau$ used by the proposed layer-by-layer iterative optimization algorithm is presented for different number of intermediate layers $L$ of the SIM architecture. Fig. \ref{Simu_Fig_perfect_56} (a) shows that the proposed layer-by-layer iterative optimization algorithm exhibits rapid convergence, since it can converge within $\tau=2$. Furthermore, Fig. \ref{Simu_Fig_perfect_56} (b) presents the average spectral efficiency $R$ versus the number of iterations $\tau$ in our proposed layer-by-layer iterative optimization algorithm. The results are parameterized by the number of reconfigurable elements $N$, with the number of intermediate layers set to $L=3$. Although the convergence speed is reduced as the number of reconfigurable elements $N$ increases, it achieves convergence within 5 iterations.

\begin{figure}[!t]
    \centering
    \subfloat[Ideal SIM hardware with $\sigma_\mathrm{p}^2=0$.]
    {\begin{minipage}{1\linewidth}
        \centering
        \includegraphics[width=2.8in]{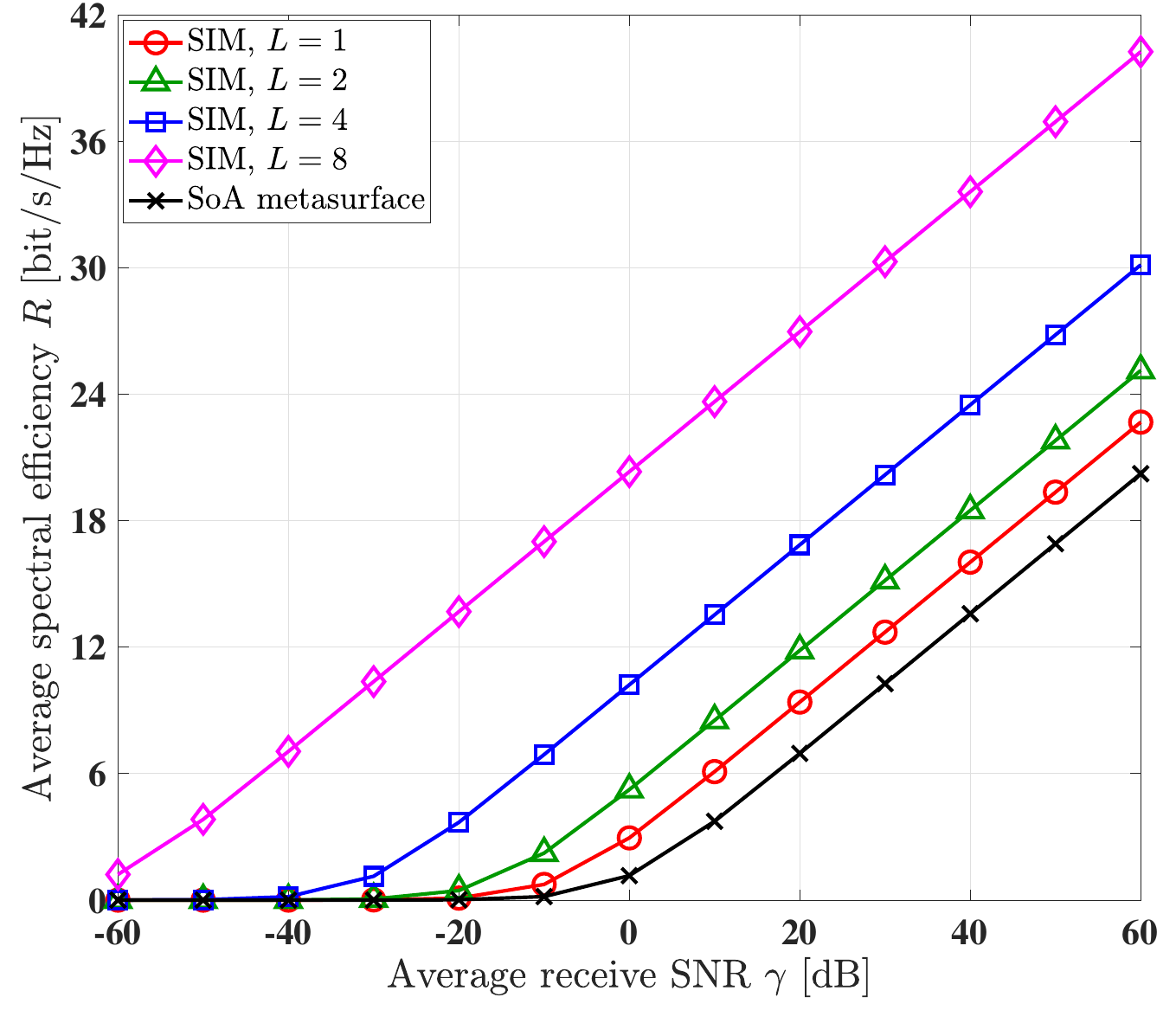}
    \end{minipage}}\\
    \subfloat[Von-Mises distribution with $\sigma_\mathrm{p}^2=10^{-2}$.]
    {\begin{minipage}{1\linewidth}
        \centering
        \includegraphics[width=2.8in]{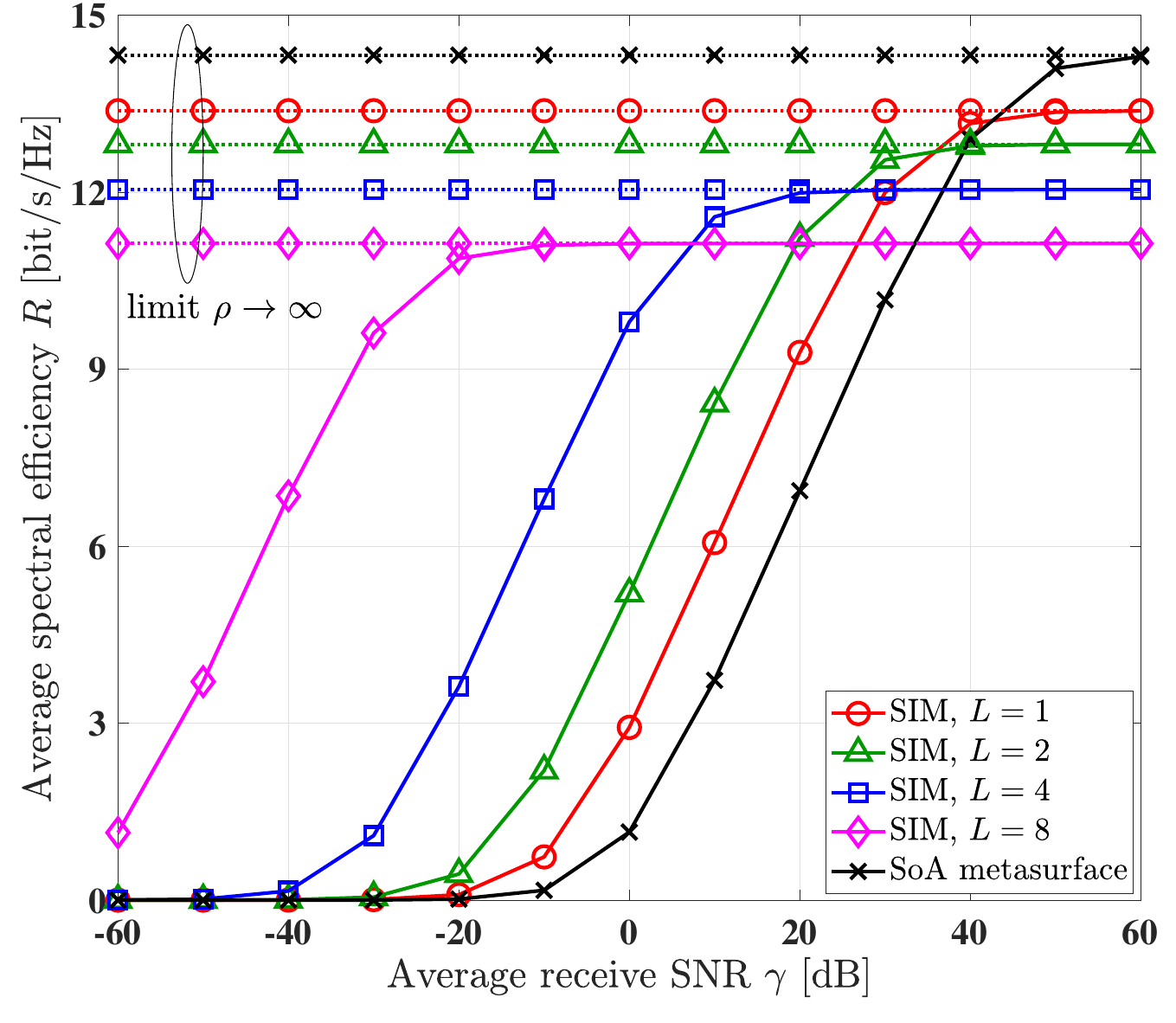}
    \end{minipage}}\\
    \subfloat[Uniform distribution with $\sigma_\mathrm{p}^2=10^{-1}$.]
    {\begin{minipage}{1\linewidth}
        \centering
        \includegraphics[width=2.8in]{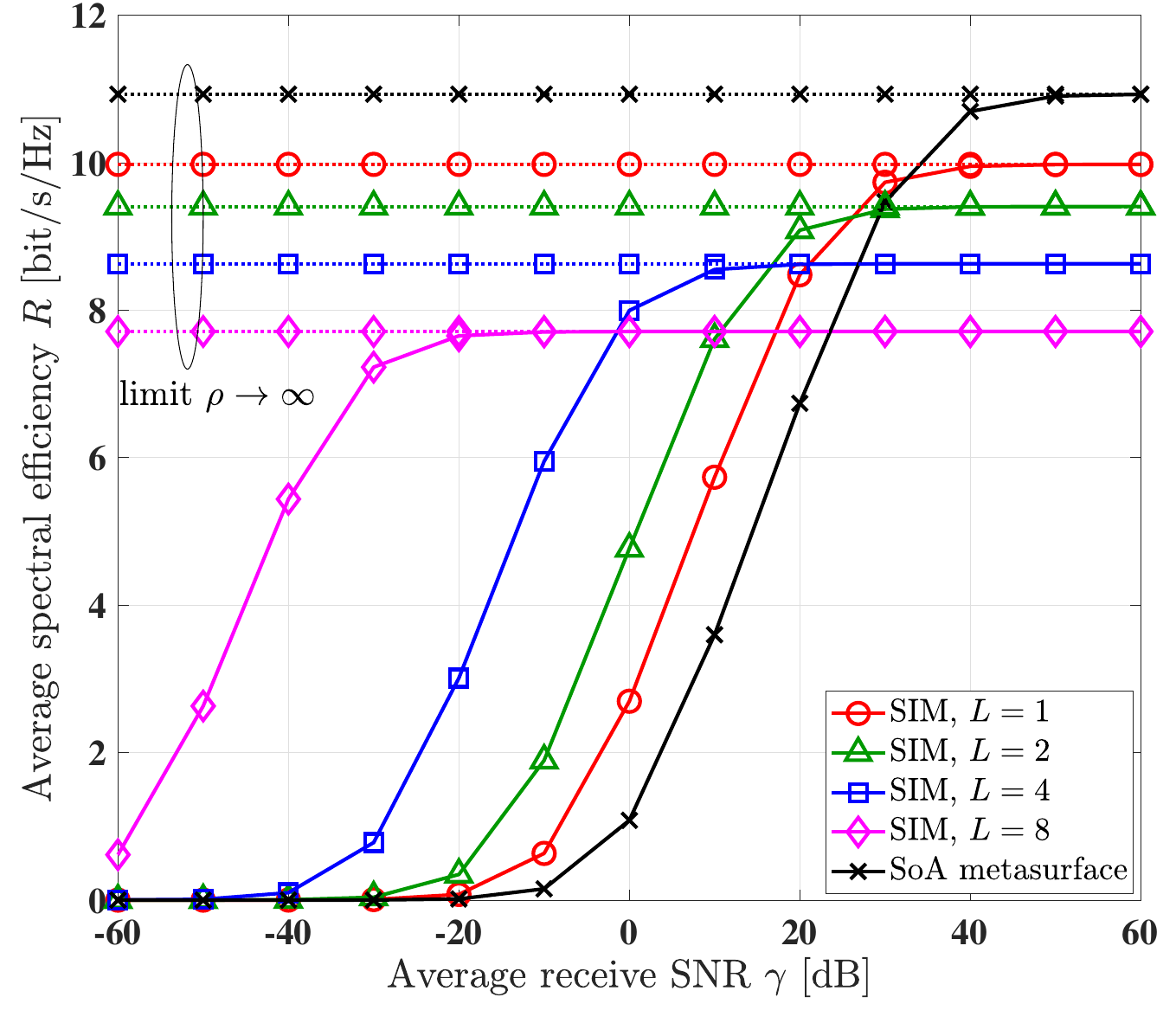}
    \end{minipage}}
    \caption{The average spectral efficiency $R$ versus the average receive SNR $\gamma$ in the SIM, with different values of phase tuning error variance $\sigma_\mathrm{p}^2$.}\label{Simu_Fig_imperfect_123}
\end{figure}

\subsection{Effect of Metasurface Phase Tuning Error}
Fig. \ref{Simu_Fig_imperfect_123} investigates the effect of the metasurface phase tuning error on the SIM-based transceiver. Specifically, Fig. \ref{Simu_Fig_imperfect_123} (a) shows that the spectral efficiency can be improved upon increasing the average receive SNR when the reconfigurable elements have no phase tuning errors, i.e., the phase tuning error variance $\sigma_\mathrm{p}^2=0$. However, when the phase tuning error variance is $\sigma_\mathrm{p}^2>0$, the average spectral efficiency saturates in the high-SNR region. Moreover, Fig. \ref{Simu_Fig_imperfect_123} (b) and Fig. \ref{Simu_Fig_imperfect_123} (c) show that in the low-SNR region, employing more intermediate layers for the SIM is beneficial for improving the spectral efficiency. This is due to the fact that in the low-SNR region, the signal received by the user is mainly contaminated by the additive noise and employing more intermediate layers is useful for increasing the beamforming gain. By contrast, at high SNR, the conventional SoA metasurface outperforms the proposed SIM. This is especially true when phase noise is present at the SIM, as the metasurface phase tuning error becomes the dominant factor degrading the spectral efficiency. Furthermore, the signal distortion resulting from the phase tuning error becomes more severe as the number of SIM layers increases, since each layer introduces additional phase noise, further attenuating the signal. Therefore, the number of intermediate layers in the SIM architecture should be chosen based on both the channel environment and the metasurface hardware quality. More intermediate layers can be employed in low-SNR scenarios, provided that the hardware quality is high enough to mitigate the effects of phase noise and signal distortion.

\begin{figure}[!t]
    \centering
    \includegraphics[width=2.8in]{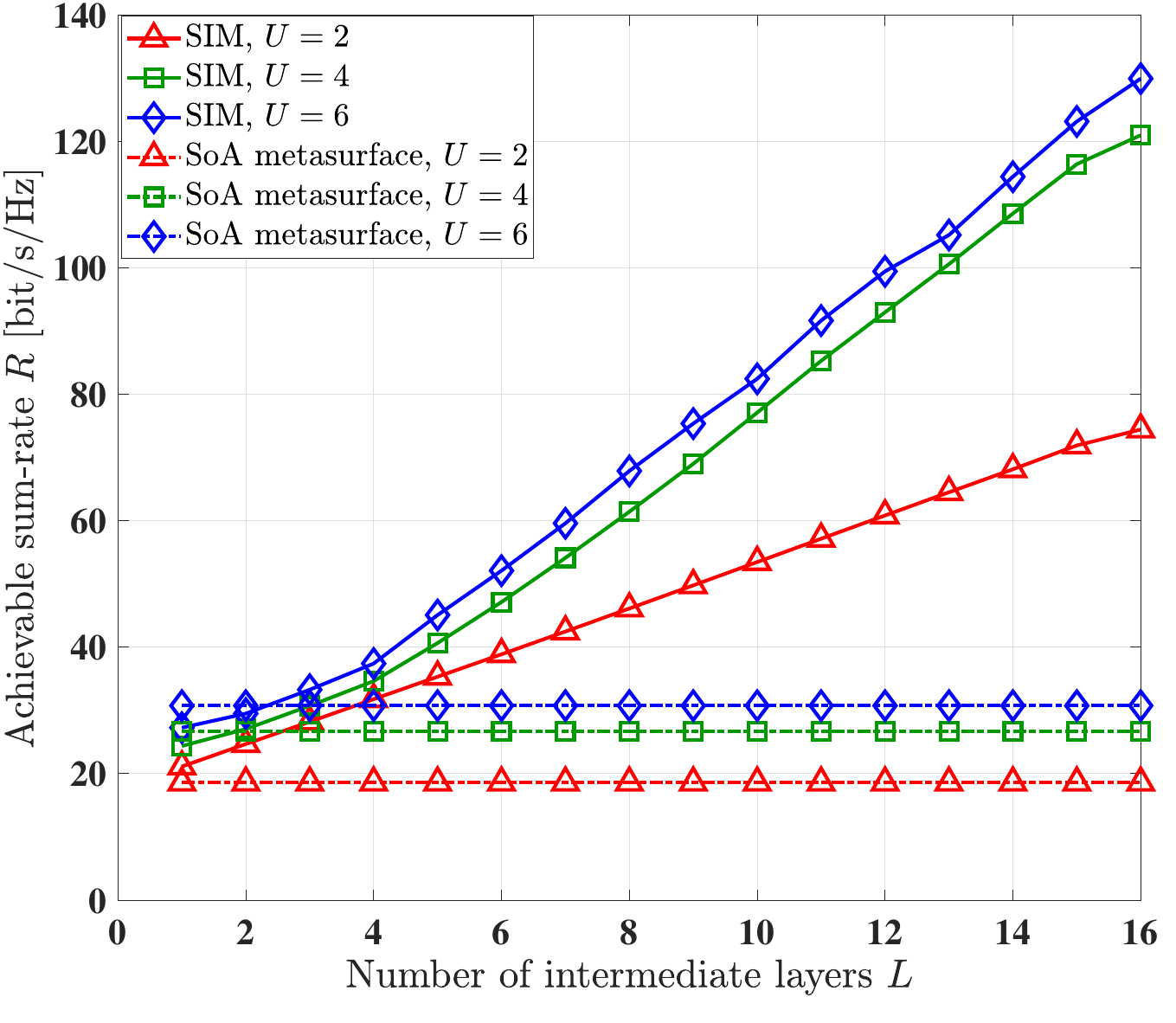}
    \caption{The achievable sum-rate $R$ versus the number of the intermediate layers $L$ in the SIM architecture, with different numbers of UEs $U$.}\label{Simu_Fig_multi_1}
\end{figure}

\begin{figure}[!t]
    \centering
    \includegraphics[width=2.8in]{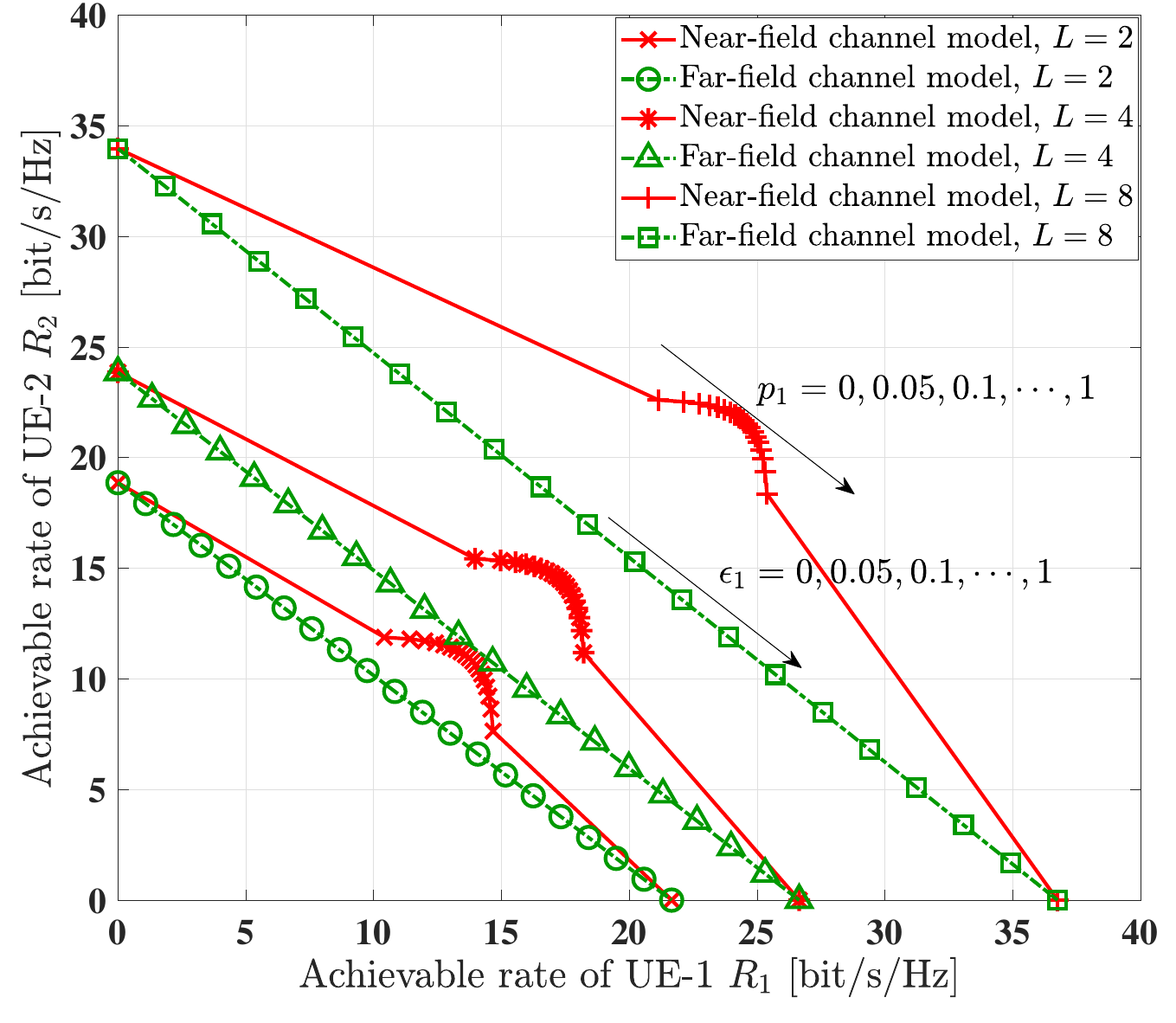}
    \caption{Performance comparison of the achievable rate in the near-field channel model and the far-field channel model.}\label{Simu_Fig_multi_2}
\end{figure}

\subsection{Achievable Rate Comparison in Near-field and Far-field}
Next, we focus on a multi-user scenario. Referring to \cite{wu2023multiple}, in the far-field channel model, only the users at different angles can be simultaneously supported, since the EM field is modeled by plane waves and the multi-user access relies on the angular orthogonality to distinguish multiple users at different angles. By contrast, thanks to the spherical wave-front characteristics in the near-field channel model, the spatial resources in both the angle-domain and the distance-domain can be exploited for supporting multiple users~\cite{an2023toward_beamfocusing}.

Fig. \ref{Simu_Fig_multi_1} compares the achievable sum-rate $R$ versus the number of intermediate layers $L$ in the SIM architecture supporting different numbers of UEs $U$, where $U=2$ means that UE-1 and UE-2 are supported, $U=4$ means that UE-1, UE-2, UE-3 and UE-4 are supported, while $U=6$ means that all 6 users seen in Fig. \ref{Fig_system_setup} are supported. Fig. \ref{Simu_Fig_multi_1} shows that higher sum-rate can be achieved by supporting multiple users. Furthermore, with the increase of the number of the intermediate layers, the achievable sum-rate of the SIM is higher than that of the SoA metasurface.

Moreover, in Fig. \ref{Simu_Fig_multi_2} the SIM-aided BS supports a pair of users in the same spatial angle to explicitly examine the performance comparison of the near-field channel model and the far-field channel model. Specifically, the SIM-aided BS supports the first pair of users, i.e., UE-1 and UE-2. Since UE-1 and UE-2 are located in the same spatial angle, the far-field channel model fails to support UE-1 and UE-2 simultaneously. The two significant jumps in the curves for the near-field channel model, corresponding to zero UE-1 rate and zero UE-2 rate, occur because of unfair power allocation between the two users. At these unfiar points, where all power is allocated to one user, the other user receives insufficient signal strength, causing their rate to drop to zero. Therefore, we employ the conventional time-division multiple access (TDMA) scheme for the far-field channel model. Specifically, the achievable sum-rate in the TDMA scheme can be presented as $R=\epsilon_1R_1+\epsilon_2R_2$, where $\epsilon_1$ and $\epsilon_2$ denote the orthogonal time resource ratios allocated for UE-1 and UE-2, respectively, satisfying $\epsilon_1+\epsilon_2=1$. By contrast, in the near-field channel model, since the spatial distance domain can be exploited, the SIM-aided BS can support UE-1 and UE-2 simultaneously. Fig. \ref{Simu_Fig_multi_2} shows that the achievable sum-rate of the near-field channel model is higher than that of the far-field channel model, especially upon increasing of the number of intermediate layers. This demonstrates that the near-field model can support both users simultaneously by exploiting spatial separation, whereas in unfair power allocation scenarios, the lack of power for one of the users causes the rate to become zero.

\section{Conclusions}\label{Conclusion}
A novel SIM-based hybrid beamforming paradigm was proposed for wideband wireless communication systems, where the spectral efficiency maximization problem was formulated based on the optimization of both the holographic beamformer and of the digital beamformer. Specifically, in the holographic beamformer, we proposed a layer-by-layer iterative optimization algorithm for maximizing the sum of the baseband eigen-channel gains of all users, which was achieved by optimizing the phase shift of the reconfigurable elements in each layer alternately. The MMSE criterion was employed for the digital beamformer to mitigate the inter-user interference. Moreover, the mitigation of the SIM phase tuning error was also considered in the beamforming design by exploiting its statistics. Additionally, the theoretical analysis of our SIM-based hybrid beamforming design was derived. We showed that the spectral efficiency saturates in the high-SNR region due to the phase tuning errors. Our simulation results unveiled the following insights. Firstly, they demonstrated that the SIM-aided holographic beamformer achieves higher spectral efficiency than the SoA single-layer metasurface aided holographic beamformer. Furthermore, the metasurface phase tuning errors have a critical effect on the spectral efficiency especially in the high-SNR region. Hence, the number of SIM layers should be carefully designed to maximize the spectral efficiency according to the specific hardware quality and the propagation environment. Finally, we showed that the near-field channel is capable of supporting multiple users by fully exploiting spatial resources in both the angle-domain and the distance-domain.

\appendices
\section{Proof of Theorem \ref{Theorem_1}}\label{Appendix_A}
Firstly, we focus on the SIM phase tuning error $\widetilde{\theta}_n^{(l)}$ obeying the uniform distribution, i.e., $\widetilde{\theta}_n^{(l)}\sim\mathcal{U}(-\iota_\mathrm{p},\iota_\mathrm{p})$. When $\widetilde{\theta}_n^{(l)}\sim\mathcal{U}(-\iota_\mathrm{p},\iota_\mathrm{p})$, the $i$th-order moment $\mathbb{E}[\widetilde{\theta}_n^{(l)i}]$ of $\widetilde{\theta}_n^{(l)}$ is equal to 0 when $i$ is odd and equal to $\frac{1}{i+1}\iota_\mathrm{p}^i$ when $i$ is even. Thus, we arrive at
\begin{align}\label{Appendix_A_1}
    \notag\mathbb{E}\left[\mathrm{e}^{j\widetilde{\theta}_n^{(l)}}\right]
    =&\sum_{i=0}^{\infty}\frac{(-1)^i}{(2i)!}\mathbb{E}
    \left[\widetilde{\theta}_n^{(l)2i}\right]
    =\sum_{i=0}^{\infty}\frac{(-1)^i\iota_\mathrm{p}^{2i}}{(2i+1)!}
    =\frac{\sin(\iota_\mathrm{p})}{\iota_\mathrm{p}}\\
    =&\xi.
\end{align}
Hence the SIM phase tuning error variance is $\sigma_\mathrm{p}^2=\mathbb{E}[\widetilde{\theta}_n^{(l)2}]
=\frac{1}{3}\iota_\mathrm{p}^2$. Then, we assume that the SIM phase tuning error $\widetilde{\theta}_n^{(l)}$ follows the von-Mises distribution, i.e., $\widetilde{\theta}_n^{(l)}\sim\mathcal{VM}(0,\varpi_\text{p})$, which satisfies~\cite{hillen2017moments}
\begin{align}\label{Appendix_A_2}
    \mathbb{E}\left[\mathrm{e}^{j\widetilde{\theta}_n^{(l)}}\right]=\frac{I_{1}(\varpi_\text{p})}
    {I_{0}(\varpi_\text{p})}=\xi.
\end{align}
Therefore, the SIM phase tuning error variance becomes $\sigma_\mathrm{p}^2=\mathbb{E}[\widetilde{\theta}_n^{(l)2}]=\frac{1}{\varpi_\text{p}}$.

According to (\ref{System_Model_6_3}), (\ref{Beamforming_Design_2}), (\ref{Appendix_A_1}) and (\ref{Appendix_A_2}), we have
\begin{align}\label{Appendix_A_3}
    \notag\mathbf{h}_u^{(k)\mathrm{H}}
    =&\mathbb{E}\left[\mathbf{g}_u^{(k)\mathrm{H}}\mathbf{\Theta}^{(L)}\mathbf{F}_k^{(L)}\cdots
    \mathbf{\Theta}^{(1)}\mathbf{F}_k^{(1)}\mathbf{\Theta}^{(0)}\mathbf{F}_k^{(0)}\right]\\
    \notag=&\mathbf{g}_u^{(k)\mathrm{H}}
    \overline{\mathbf{\Theta}}^{(L)}\mathbf{F}_k^{(L)}
    \cdots\overline{\mathbf{\Theta}}^{(1)}\mathbf{F}_k^{(1)}
    \overline{\mathbf{\Theta}}^{(0)}\mathbf{F}_k^{(0)}\cdot\\
    \notag&\mathbb{E}\left[\widetilde{\mathbf{\Theta}}^{(L)}\right]\cdots
    \mathbb{E}\left[\widetilde{\mathbf{\Theta}}^{(1)}\right]
    \mathbb{E}\left[\widetilde{\mathbf{\Theta}}^{(0)}\right]\\
    =&\xi^{L+1}\mathbf{g}_u^{(k)\mathrm{H}}\overline{\mathbf{A}}^{(k)}.
\end{align}
Furthermore, according to (\ref{System_Model_6_3}), (\ref{Beamforming_Design_3}), (\ref{Appendix_A_1}) and (\ref{Appendix_A_2}), $\mathbb{E}[\mathbf{\Theta}^{(L)\mathrm{H}}\mathbf{g}_u^{(k)}\mathbf{g}_u^{(k)\mathrm{H}}
\mathbf{\Theta}^{(L)}]$ can be formulated as $\mathbb{E}[\mathbf{\Theta}^{(L)\mathrm{H}}
\mathbf{g}_u^{(k)}\mathbf{g}_u^{(k)\mathrm{H}}\mathbf{\Theta}^{(L)}]
=\xi^2\overline{\mathbf{\Theta}}^{(L)\mathrm{H}}\mathbf{g}_u^{(k)}\mathbf{g}_u^{(k)\mathrm{H}}
\overline{\mathbf{\Theta}}^{(L)}+(1-\xi^2)(\mathbf{g}_u^{(k)}\mathbf{g}_u^{(k)\mathrm{H}})
\odot\mathbf{I}_N=\mathbf{\Phi}_k^{(L-1)}$. Similarly, $\mathbb{E}[\mathbf{\Theta}^{(L-1)\mathrm{H}}\mathbf{F}_k^{(L)\mathrm{H}}
\mathbf{\Phi}_k^{(L)}\mathbf{F}_k^{(L)}\mathbf{\Theta}^{(L-1)}]$ can be expressed as $\mathbb{E}[\mathbf{\Theta}^{(L-1)\mathrm{H}}\mathbf{F}_k^{(L)\mathrm{H}}
\mathbf{\Phi}_k^{(L)}\mathbf{F}_k^{(L)}\mathbf{\Theta}^{(L-1)}]
=\xi^2\overline{\mathbf{\Theta}}^{(L-1)\mathrm{H}}
\mathbf{F}_k^{(L)\mathrm{H}}\mathbf{\Phi}_k^{(L)}\mathbf{F}_k^{(L)}
\overline{\mathbf{\Theta}}^{(L-1)}+(1-\xi^2)(\mathbf{F}_k^{(L)\mathrm{H}}\mathbf{\Phi}_k^{(L)}
\mathbf{F}_k^{(L)})\odot\mathbf{I}_N
=\mathbf{\Phi}_k^{(L-2)}$. By leveraging the above iterative operation, we can get $\mathbb{E}[\mathbf{h}_u^{(k)}\mathbf{h}_u^{(k)\mathrm{H}}]
=\mathbf{F}_k^{(0)\mathrm{H}}\mathbf{\Phi}_k^{(0)}\mathbf{F}_k^{(0)}$. Therefore, we can express the covariance matrix of $\widetilde{\mathbf{h}}_u^{(k)}$ as
\begin{align}\label{Appendix_A_7}
    \notag\mathbf{C}_{\widetilde{\mathbf{h}}_u^{(k)}\widetilde{\mathbf{h}}_u^{(k)}}
    =&\mathbb{E}\left[\mathbf{h}_u^{(k)}\mathbf{h}_u^{(k)\mathrm{H}}\right]-
    \mathbb{E}\left[\mathbf{h}_u^{(k)}\right]\mathbb{E}\left[\mathbf{h}_u^{(k)\mathrm{H}}\right]\\
    =&\mathbf{F}_k^{(0)\mathrm{H}}\mathbf{\Phi}_k^{(0)}\mathbf{F}_k^{(0)}
    -\xi^{2(L+1)}\overline{\mathbf{A}}^{(k)\mathrm{H}}\mathbf{g}_u^{(k)}
    \mathbf{g}_u^{(k)\mathrm{H}}\overline{\mathbf{A}}^{(k)}.
\end{align}

\section{Proof of Theorem \ref{Theorem_2}}\label{Appendix_B}
When $d_l\rightarrow 0$ for $l=0,1,\cdots,L$, we have $\beta_{n_2,n_1}^{(l)}=1$ when $n_1=n_2$ and $\beta_{n_2,n_1}^{(l)}=0$ when $n_1\neq n_2$. Therefore, the channel link spanning from the $n_1$th element in layer-$(l-1)$ to the $n_2$th element in layer-$l$ ($l=1,2,\cdots,L$) at the central carrier frequency can be represented as:
\begin{align}\label{Appendix_B_1}
    \notag\mathbf{F}_k^{(l)}=&\mathbf{Diag}
    \left\{\mathrm{e}^{-\jmath \frac{2\pi}{\lambda_k}\left\|\mathbf{p}_{1}^{(l)}
    -\mathbf{p}_{1}^{(l-1)}\right\|},
    \mathrm{e}^{-\jmath \frac{2\pi}{\lambda_k}\left\|\mathbf{p}_{2}^{(l)}
    -\mathbf{p}_{2}^{(l-1)}\right\|},\cdots,\right.\\
    &\left.\mathrm{e}^{-\jmath \frac{2\pi}{\lambda_k}\left\|\mathbf{p}_{N}^{(l)}
    -\mathbf{p}_{N}^{(l-1)}\right\|}\right\}.
\end{align}
Therefore, we can obtain
\begin{align}\label{Appendix_B_2}
    \notag&\overline{\mathbf{\Theta}}^{(L)}\mathbf{F}_k^{(L)}
     \cdots\overline{\mathbf{\Theta}}^{(1)}\mathbf{F}_k^{(1)}\overline{\mathbf{\Theta}}^{(0)}\\
    \notag=&\mathbf{Diag}
    \left\{\mathrm{e}^{-\jmath \left(\overline{\theta}_1^{(0)}
    +\sum_{l=1}^{L}\left(\overline{\theta}_1^{(l)}-\frac{2\pi}{\lambda_k}\left\|\mathbf{p}_{1}^{(l)}
    -\mathbf{p}_{1}^{(l-1)}\right\|\right)\right)},\right.\\
    \notag&\left.\mathrm{e}^{-\jmath \left(\overline{\theta}_2^{(0)}
    +\sum_{l=1}^{L}\left(\overline{\theta}_2^{(l)}-\frac{2\pi}{\lambda_k}\left\|\mathbf{p}_{2}^{(l)}
    -\mathbf{p}_{2}^{(l-1)}\right\|\right)\right)},\cdots,\right.\\
    \notag&\left.\mathrm{e}^{-\jmath \left(\overline{\theta}_N^{(0)}
    +\sum_{l=1}^{L}\left(\overline{\theta}_N^{(l)}-\frac{2\pi}{\lambda_k}\left\|\mathbf{p}_{N}^{(l)}
    -\mathbf{p}_{N}^{(l-1)}\right\|\right)\right)}\right\}\\
    =&\mathbf{\Xi}_k.
\end{align}
Upon substituting (\ref{Appendix_B_2}) into (\ref{Beamforming_Design_5}) and (\ref{Beamforming_Design_6}), we can arrive at
\begin{align}\label{Appendix_B_3}
    \overline{\mathbf{h}}_u^{(k)\mathrm{H}}=\xi^{L+1}\mathbf{g}_u^{(k)\mathrm{H}}
    \mathbf{\Xi}_k\mathbf{F}_k^{(0)},
\end{align}
and
\begin{align}\label{Appendix_B_4}
    \mathbf{C}_{\widetilde{\mathbf{h}}^{(k)}\widetilde{\mathbf{h}}^{(k)}}
    =&\left(1-\xi^{2(L+1)}\right)\mathbf{F}_k^{(0)\mathrm{H}}
    \left(\left(\mathbf{g}_u^{(k)}\mathbf{g}_u^{(k)\mathrm{H}}\right)
    \odot\mathbf{I}_N\right)\mathbf{F}_k^{(0)}.
\end{align}
Based on (\ref{Performance_Analysis_1}), (\ref{Appendix_B_2}), (\ref{Appendix_B_3}) and (\ref{Appendix_B_4}), we can arrive at (\ref{Performance_Analysis_3}).

\bibliographystyle{IEEEtran}
\bibliography{IEEEabrv,TAMS}
\end{document}